\documentclass[twocolumn]{aastex6}

\usepackage{amsmath}
\usepackage{rotating}
\usepackage{longtable}
\usepackage{float}
\usepackage{verbatim}
\usepackage[pass]{geometry}

\usepackage{longtable}
\shorttitle{Know the planet, know the star}
\shortauthors{Sandford \& Kipping}

\begin{document}

\title{Know the Planet, Know the Star: 

Precise Stellar Densities from \textit{Kepler} Transit Light Curves}

\author{Emily Sandford\altaffilmark{1}$^\star$ and David Kipping\altaffilmark{1}}
\affil{$^1$Department of Astronomy, Columbia University, 550 West 120th Street, New York, NY 10027, USA}

\email{$^\star${esandford@astro.columbia.edu}}

% SE 52 stars (without TTVs)
% TC 26 stars (without TTVs)
% MP2 16 stars (without TTVs)
% = 94 stars total
\begin{abstract}
%The \textit{Kepler} space telescope has revolutionized exoplanetary science with unprecedentedly precise photometric measurements of the light curves of transiting planets. 

The properties of a transiting planet's host star are written in its transit light curve. The light curve can reveal the stellar density ($\rho_*$) and the limb darkening profile in addition to the characteristics of the planet and its orbit. For planets with strong prior constraints on orbital eccentricity, we may measure these stellar properties directly from the light curve; this method promises to aid greatly in the characterization of transiting planet host stars targeted by the upcoming NASA TESS mission and any long-period, singly-transiting planets discovered in the same systems. Using Bayesian inference, we fit a transit model, including a nonlinear limb darkening law, to 66 \textit{Kepler} transiting planet hosts to measure their stellar properties. We present posterior distributions of $\rho_*$, limb-darkening coefficients, and other system parameters for these stars. We measure densities to within $5\%$ for the majority of our target stars, with the dominant precision-limiting factor being the signal-to-noise ratio of the transits. $95 \%$ of our measured stellar densities are in $3\sigma$ or better agreement with previously published literature values. We make posterior distributions for all of our target KOIs available online at \url{https://doi.org/10.5281/zenodo.1028515}.

\end{abstract}


\section{Introduction} \label{sec:intro}
Since its launch in 2009, the \textit{Kepler} mission has discovered over 4500 transiting exoplanet candidates, nearly 2000 of which have been validated at $> 99\%$ confidence \citep{morton16}. Furthermore, \textit{Kepler} transit light curve modeling (e.g., \citealt{batalha13}) has yielded precise constraints on the characteristics of these planet candidates and their orbits; the transit depth, for example, reveals the size of the planet relative to its host star.

Encoded in each transit light curve, however, is not just the character of the transiting planet, but also properties of the host star. In particular, the stellar density ($\rho_*$) can be derived analytically from the transit duration using Kepler's third law (\citealt{seager03}; see section~\ref{sec:theory}), provided the eccentricity of the planet's orbit is well-constrained. Furthermore, the star's limb darkening behavior influences the shape of the transit light curve during planetary ingress and egress (see e.g. \citealt{knutson07}). In other words, the star's interior and atmospheric properties manifest themselves in the shape of the transit light curve. 

By fitting transit models to the \textit{Kepler} light curves, we can measure these stellar properties very precisely. We have several motivations to measure stellar properties for a large sample of \textit{Kepler} hosts in this way. First, transit modeling serves as an independent check on other means of measuring stellar properties. In the case of $\rho_*$, such methods include asteroseismology, as well as spectroscopy plus isochrone modeling. These methods rest on different assumptions and, often, different input data.

%We may compare transit-derived stellar properties to the theoretical predictions of stellar atmosphere modeling.

Transit modeling also offers an independent test of stellar atmosphere theory, particularly with regard to limb darkening behavior. Such behavior is usually expressed in the form of an analytic stellar intensity profile weighted by limb-darkening coefficients (LDCs). Traditionally, LDCs are adopted from the theoretical predictions of stellar atmosphere modeling codes (see e.g. \citealt{sing10, claret00}). This practice is known to introduce biases in exoplanet parameters subsequently derived from the light curve \citep{espinoza15}. Measuring the LDCs directly from the light curve enables an empirical check of these stellar atmosphere model predictions.

Furthermore, the commonly used quadratic limb darkening law, which has two LDCs, is known to be less accurate than laws with three or four LDCs \citep{kipping16ldc, sing10}. By adopting a three-parameter nonlinear limb darkening law in transit light curve modeling and building up an empirical catalog of the fitted LDCs, we may address some of these inaccuracies.

We may also use transit fitting to derive stellar properties for stars that are not amenable to traditional analysis. For example, asteroseismology, which yields extremely precise constraints on $\rho_*$ (typical fractional uncertainties $\lesssim 5\%$), is only possible for stars which are bright (Kepler-band magnitude $\lesssim 12$) and massive  ($\gtrsim 1 \textrm{M}_{\odot}$) \citep{huber13}. Most stars are smaller and dimmer than this. 

Finally, measuring stellar properties from transit light curves alone allows us to characterize planet-hosting stars without committing telescope time to obtain follow-up observations. In an era of large-scale surveys of transiting exoplanets, such efficiency will be crucial. The NASA Transiting Exoplanet Survey Satellite (TESS; \citealt{ricker14}), scheduled to launch in 2018, is expected to discover thousands of transiting planets orbiting stars observed at two-minute cadence, but potentially tens of thousands more around other stars in its field of view \citep{sullivan15}. The Large Synoptic Survey Telescope (LSST; \citealt{lsst}), expected to begin full-scale science operations in 2023, will discover thousands more. 

In this work, we fit transit models to a large sample of \textit{Kepler} host stars to build an empirical catalog of transit-derived stellar densities and limb darkening coefficients and demonstrate that this method is capable of delivering precise constraints on these stellar parameters. In Section~\ref{sec:methods}, we describe our host star target selection and detail our data analysis, including data processing, detrending, and Markov chain Monte Carlo (MCMC) approach to fitting the transit model. In Section~\ref{sec:results}, we present results of this analysis, including the full posterior distributions of the stellar density and LDCs. We specifically discuss the types of planet-star systems for which this method succeeds in producing high-precision constraints on stellar density in Section~\ref{sec:discussion}. We conclude and highlight this approach's potential to aid in the characterization of singly-transiting planets discovered by the upcoming NASA TESS mission in Section~\ref{sec:conc}.

\section{Methods} \label{sec:methods}
\subsection{How to measure $\rho_*$ from a transit light curve}\label{sec:theory}

\begin{figure}
\begin{centering}
\includegraphics[width=0.45\textwidth]{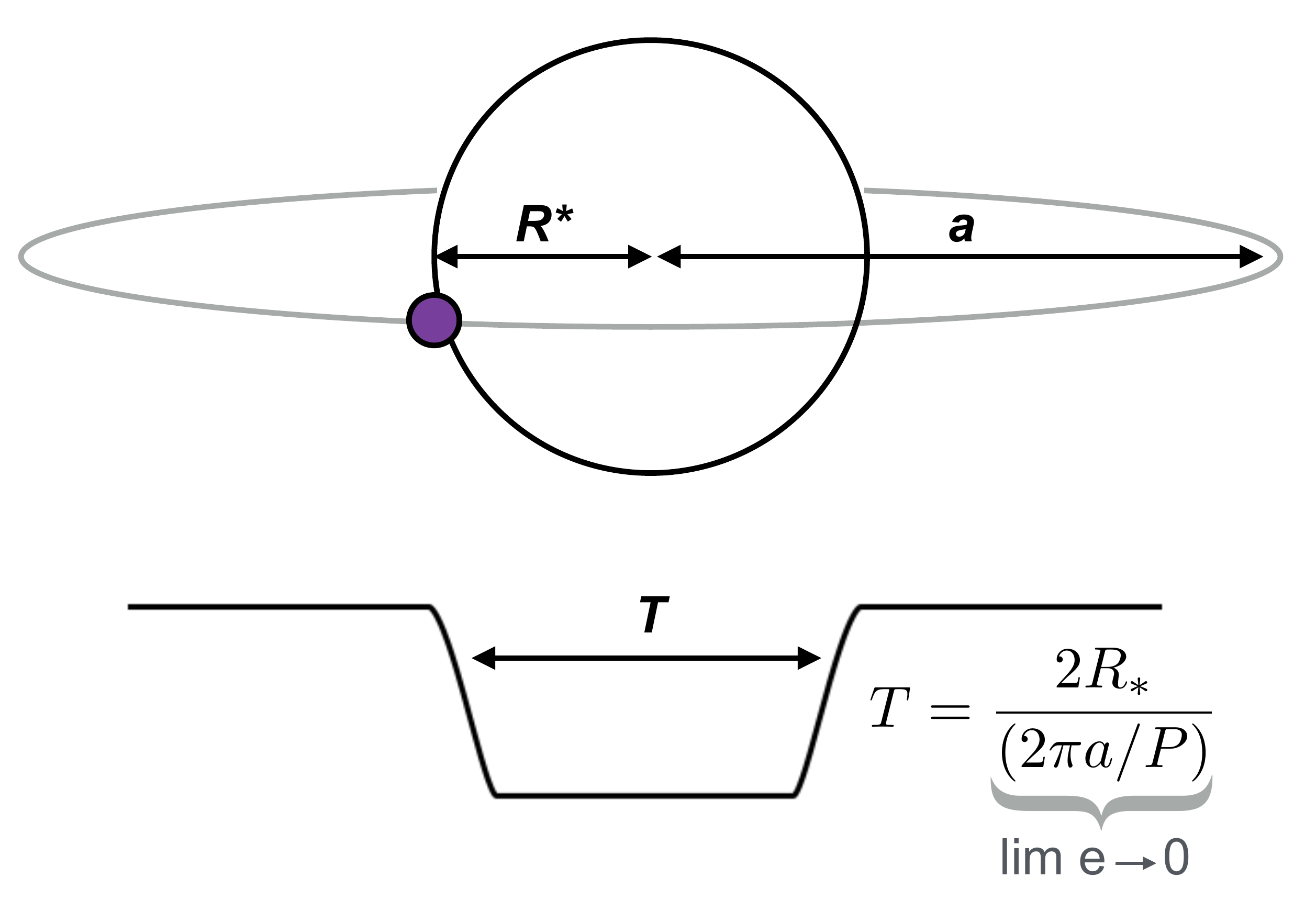}
\caption{The transit duration $T$ is equal to the stellar diameter divided by the mean orbital velocity, which is equal to $2\pi a/P$ in the case of a circular orbit. Rearrangement of the equation in the lower panel yields the normalized semimajor axis $a/R_*$. An analogous calculation is possible for planets on eccentric orbits (for which orbital velocity varies with phase), provided the eccentricity is known.}
\label{fig:aR}
\end{centering}
\end{figure}

\cite{seager03} demonstrated that the mean stellar density $\rho_*$ can be measured from a transit light curve without any direct measurement of the stellar mass $M_*$ or radius $R_*$ as a result of Kepler's third law. Figure~\ref{fig:aR} offers some intuition about this procedure in the case of a circular orbit, and we sketch the analytic derivation of the circular-orbit case here.

We begin with Kepler's third law:

\begin{equation}
    \frac{P^2}{4\pi^2} = \frac{a^3}{G(M_* + M_p)} \simeq \frac{a^3}{GM_*},
\end{equation}

where the right-hand side assumes that $M_p \ll M_*$. Dividing and multiplying the right-hand side of this equation by the stellar volume, $\frac{4}{3}\pi R_*^3$, we obtain:

\begin{equation}
    \frac{P^2}{4\pi^2} = \frac{3(a/R_*)^3}{4\pi G \rho_*}
\end{equation}

Rearrangement yields:

\begin{equation}\label{eq:rhostar}
    \rho_* = \frac{3\pi (a/R_*)^3}{G P^2}
\end{equation}

Therefore, to measure $\rho_*$, we need only know the orbital period $P$ and normalized semimajor axis $a/R_*$ of a planet orbiting the star. (In particular, neither $M_*$ nor $R_*$ is necessary to obtain $\rho_*$.) Both $P$ and $a/R_*$ are directly measurable from the transit light curve: $P$ is the interval between successive transits, and $a/R_*$ can be derived from the transit duration. In the case of a circular orbit, $a/R_*$ follows trivially from the transit duration and $P$ (see Figure~\ref{fig:aR}):

\begin{equation}
    T = \frac{2R_*}{(2\pi a / P)}
\end{equation}

Rearrangement of this equation yields the normalized semimajor axis $a/R_*$:

\begin{equation}
    \frac{a}{R_*} = \frac{P}{\pi T}
\end{equation}

\twocolumngrid

However, in general, the eccentricity $e$ of the transiting planet's orbit also influences the transit duration $T$. The exact solution for $T$ in the case of an eccentric orbit involves solving a quartic equation in $\cos{f}$, where $f$ is the true anomaly (see \citealt{kipping08,kipping10} for details). However, \cite{kipping10} found the following approximate expression for $T$ under the simplifying assumption that the planet-star separation does not change during the transit:

\begin{equation}
    T \simeq \frac{P}{\pi} \frac{\varrho_c^2}{\sqrt{1 - e^2}}\arcsin{\left( \frac{\sqrt{1-(a/R_*)^2\varrho_c^2\cos^2{i}}}{(a/R_*)\varrho_c\sin{i}} \right)}
\end{equation}

where $\varrho_c$ is the separation between the planet and star at mid-transit, in units of stellar radii.

Since $e$ and $\rho_*$ both influence the transit duration $T$, it is necessary to have a precise constraint on the $e$ in order to derive a precise constraint on $\rho_*$ \citep{kipping10}. For some planets, such as planets with observed secondary eclipses, $e$ is directly measurable (e.g. \citealt{knutson07nat}); for others, such as planets on very short-period orbits which are expected to tidally circularize quickly or planets in compact multi-planet systems, dynamical stability constrains $e$ to low values. For each of these categories of planet---secondary-eclipse planets, tidally circularized planets, and multi-planet systems---we may express the existing eccentricity constraint as a Bayesian prior on $e$. In sections~\ref{subsubsec:SE}-\ref{subsubsec:MP}, we describe how we select a sample of \textit{Kepler} Objects of Interest (KOIs) belonging to each category for transit modeling.

We note that, in principle, $\rho_*$ could also be measured from the transits of planets with radial velocity-measured eccentricities. However, analyzing such planets requires jointly fitting the radial velocity curves, including accurate treatment of stellar activity effects. This is beyond the scope of the present study, and we defer analysis of planets with radial velocity-measured $e$ to later work. 

Assuming, then, that we have a strong $e$ prior, all we must do to measure $\rho_*$ from a transit is fit a transit model, comprising ten parameters: the transit epoch $t_0$, the orbital period $P$, the impact parameter $b$, the stellar density $\rho_*$, the ratio-of-radii $R_p/R_*$, the orbital eccentricity $e$, the argument of periastron $\omega$, and three coefficients of a modified nonlinear limb darkening law (transformed to allow for efficient sampling as described in \citealt{kipping16ldc}), $\alpha_r, \alpha_h,$ and $\alpha_\theta$. In other words, we must explore this ten-dimensional parameter space and find a region that matches the \textit{Kepler} transit data.

We use the transit-modeling code \texttt{BATMAN} \citep{kreidberg15} to compute the light curve of a given set of ten transit model parameters, compare this model to the \textit{Kepler} data, and evaluate the likelihood of the parameters. We step through the ten-dimensional parameter space and derive posterior distributions for the model parameters with \texttt{emcee} \citep{dfm13}, an affine-invariant ensemble Markov chain Monte Carlo (MCMC) sampler. Details of this procedure are given in section~\ref{subsec:modeling}.

\subsection{Sample selection}\label{subsec:targetlist}

\begin{figure*}[!ht]
\begin{center}
\includegraphics[width=0.9\textwidth]{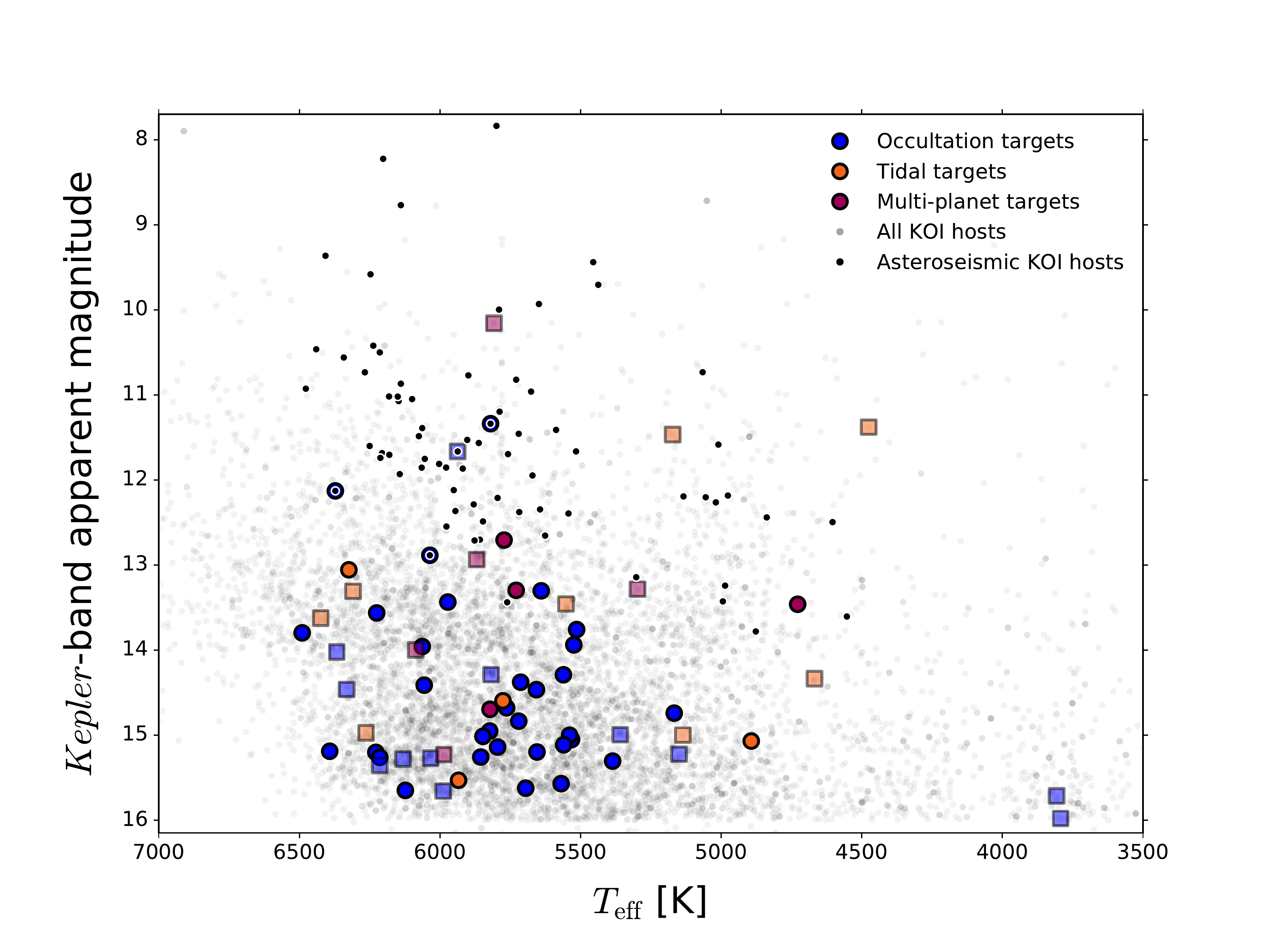}
\caption{The distribution of our target stars, compared to all KOI-hosting stars and KOI-hosting stars with asteroseismic density measurements, in $T_{\mathrm{eff}}$-magnitude space. Opaque circles represent stars for which we achieve comparable $\rho_*$ precision to asteroseismology (fractional uncertainty $\leq 5\%$); transparent squares represent stars for which we do not. Four of our targets overlap with the \cite{huber13} asteroseismic sample; see Figure~\ref{fig:AScomp} for details.}
\label{fig:target_selection}
\end{center}
\end{figure*}

Here, we describe how we select a sample of KOIs with strong eccentricity priors for transit modeling. We furthermore select the host stars of these KOIs to span a broad range in \textit{Kepler}-band magnitude and effective temperature, as shown in Figure~\ref{fig:target_selection}, in order to investigate the efficacy of this method across a wide range of stellar types.

\subsubsection{Secondary eclipse targets}\label{subsubsec:SE}

Certain exceptional transiting planets are bright enough relative to their host star that the flux from the planet-star system drops perceptibly when the planet passes behind the star. Such planets are thus detected both when they pass in front of their host stars (transit) and when they pass behind (occultation, or secondary eclipse). Clocking the planet at two points in its orbit, rather than just at transit, makes it possible to precisely constrain the eccentricity and argument of periastron ($\omega$) of the orbit (e.g., \citealt{winn10}); in other words, it places a strong prior on $e$, which allows us to measure $\rho_*$ from the planet's transit.

More specifically, we may derive constraints on $e\cos{\omega}$ and $e\sin{\omega}$ by measuring the time elapsed between mid-transit and the subsequent mid-occultation ($\Delta t$) and the relative duration of the transit compared to the occultation, $T_{\mathrm{transit}}/T_{\mathrm{occultation}}$. Approximate expressions for these constraints may be found in \cite{winn10}:

\begin{equation}
    e\cos{\omega} \simeq \frac{\pi}{4}\left( \frac{2\Delta t}{P} - 1 \right)
\end{equation}

\begin{equation}
    e\sin{\omega} \simeq \frac{T_{\mathrm{transit}}}{T_{\mathrm{occultation}}} - 1
\end{equation}

We draw KOIs with observed secondary eclipses (hereafter, ``occultation targets") from catalogs compiled by \cite{clm12} and \cite{shabram16}. \cite{shabram16} measured the eccentricities of 50 KOIs with detected secondary eclipses. Of these, five (KOI-774.01, KOI-805.01, KOI-895.01, KOI-1227.01, and KOI-1391.01) were subsequently identified as false positives in the NASA Exoplanet Archive\footnote{\url{https://exoplanetarchive.ipac.caltech.edu/}, accessed 1 August 2017.} (\citealt{akeson15}, hereafter ``NEA"), leaving 45 targets with measured eccentricities. To this list, we add a further 10 KOIs detected in secondary eclipse by \cite{clm12}. Of these, four (KOIs 1.01, 5.01, 10.01, and 412.01) are not counted as significant SE detections by \citealt{clm12}, so we advise caution in adopting our transit parameter posteriors for these targets.

We remove one of these targets (KOI-203.01) from the list due to stroboscopic starspot activity \citep{desert11} and another four (KOI-202.01, KOI-760.01, KOI-883.01, and KOI-1781.01) due to detected transit timing variations (TTVs; \citealt{holczer16}). Modeling the transits of a planet with detected TTVs is prohibitively computationally expensive, because it requires adding a new model parameter to describe every successive interval between transits \citep{teachey17}, and some of our target planets undergo hundreds of transits over \textit{Kepler}'s 4-year observational baseline.

Finally, we remove 6 occultation targets because their MCMC analysis was prohibitively slow (see Section~\ref{subsec:modeling}) as a result of their unusually high number of data points ($\sim 10^5-10^6$, compared to $\sim 10^3-10^5$ for successfully analyzed occultation targets). The resulting occultation target list, comprising 44 KOIs (the majority of our targets), is presented in Table~\ref{table:SE}.

\subsubsection{Tidally circularized targets}\label{subsubsec:TC}

Our second target population consists of KOIs with short tidal circularization timescales $\tau_{circ}$ (``tidal targets"). In general, we expect such KOIs to have approximately circular orbits ($e$ close to 0); more precisely, \cite{wang11} found that the $e$ distribution for single-planet systems with short $\tau_{circ}$ is consistent with an exponential distribution, $P(e, \lambda) = \frac{1}{\lambda}\exp{\frac{-e}{\lambda}}$, with scale parameter $\lambda = 0.00796$. Similarly, \cite{kipping13} found strong evidence that the short-period sample of RV-observe planets reside on less elliptical orbits than their longer-period counterparts, at a confidence of $11.6\sigma$.

%\pagebreak
%\begin{center}
\tabcolsep=0.015cm
\tiny
%\begin{adjustbox}{width=1.2\textwidth,center}
%\begin{longtable*}[tp]{lllllllllll}
\begin{longtable*}[tp]{@{\extracolsep{\fill}}*{11}{l}@{}}

\caption{Occultation targets, with measured transit parameters, stellar densities, and limb darkening coefficients. A machine-readable version of this table with more significant figures and the complete target list is available online. \label{table:SE}}\\

%KOI & KIC & $t_0$ [BKJD] & $P$ [days] & $b$ & $\log_{10}(\rho_*$) & $R_p/R_*$ & $\sqrt{e}\cos{\omega}$ & $\sqrt{e}\sin{\omega}$ & $\alpha_r$ & $\alpha_h$ & $\alpha_{\theta}$\\

KOI & $t_0$ [BKJD] & $P$ [days] & $b$ & $\log_{10}(\rho_*$) & $R_p/R_*$ & $\sqrt{e}\cos{\omega}$ & $\sqrt{e}\sin{\omega}$ & $\alpha_r$ & $\alpha_h$ & $\alpha_{\theta}$\\
 & &  &  & $[\mathrm{kg}/\mathrm{m}^3]$ & & & & & & \\
\hline
\endhead
1.01 & $122.7625^{+0.0003}_{-0.0006}$ & $2.47061338^{+1e-08}_{-1e-08}$ & $0.79^{+0.01}_{-0.01}$ & $3.3^{+0.07}_{-0.03}$ & $0.129^{+0.001}_{-0.001}$ & $-0.4^{+0.1}_{-0.1}$ & $-0.21^{+0.05}_{-0.08}$ & $0.01^{+0.07}_{-0.008}$ & $0.78^{+0.05}_{-0.06}$ & $0.46^{+0.08}_{-0.31}$\\
5.01 & $132.972^{+0.001}_{-0.002}$ & $4.780332^{+2e-06}_{-2e-06}$ & $1.0^{+0.7}_{-0.3}$ & $2.3^{+1.0}_{-0.3}$ & $0.07^{+0.68}_{-0.05}$ & $-0.1^{+0.2}_{-0.1}$ & $-0.15^{+0.21}_{-0.07}$ & $0.6^{+0.3}_{-0.4}$ & $0.2^{+0.7}_{-0.2}$ & $0.3^{+0.6}_{-0.3}$\\
10.01 & $121.1195^{+0.0001}_{-0.0001}$ & $3.5224985^{+1e-07}_{-1e-07}$ & $0.72^{+0.01}_{-0.03}$ & $2.68^{+0.05}_{-0.02}$ & $0.0981^{+0.0003}_{-0.0004}$ & $0.01^{+0.09}_{-0.09}$ & $-0.1^{+0.1}_{-0.1}$ & $0.1^{+0.01}_{-0.02}$ & $0.9^{+0.08}_{-0.16}$ & $0.39^{+0.05}_{-0.04}$\\
%17.01 & $121.48659^{+2e-05}_{-3e-05}$ & $3.23469935^{+8e-08}_{-8e-08}$ & $0.15^{+0.07}_{-0.03}$ & $2.85^{+0.03}_{-0.09}$ & $0.0937^{+0.0006}_{-0.0002}$ & $0.0^{+0.1}_{-0.2}$ & $0.1^{+0.1}_{-0.2}$ & $0.56^{+0.05}_{-0.02}$ & $0.8^{+0.1}_{-0.3}$ & $0.594^{+0.028}_{-0.007}$\\
%22.01 & $177.25002^{+6e-05}_{-6e-05}$ & $7.8914479^{+3e-07}_{-3e-07}$ & $0.47^{+0.01}_{-0.02}$ & $2.89^{+0.05}_{-0.03}$ & $0.0967^{+0.0002}_{-0.0002}$ & $0.0^{+0.1}_{-0.2}$ & $-0.1^{+0.2}_{-0.1}$ & $0.94^{+0.05}_{-0.11}$ & $0.33^{+0.06}_{-0.03}$ & $0.647^{+0.005}_{-0.01}$\\
%97.01 & $134.27692^{+0.00012}_{-9e-05}$ & $4.8854886^{+2e-07}_{-2e-07}$ & $0.53^{+0.02}_{-0.02}$ & $2.42^{+0.08}_{-0.05}$ & $0.083^{+0.0002}_{-0.0002}$ & $0.01^{+0.08}_{-0.09}$ & $-0.2^{+0.2}_{-0.1}$ & $0.6^{+0.2}_{-0.2}$ & $0.3^{+0.11}_{-0.06}$ & $0.63^{+0.02}_{-0.04}$\\
%98.01 & $138.0881^{+0.0003}_{-0.0006}$ & $6.7901219^{+7e-07}_{-5e-07}$ & $0.61^{+0.04}_{-0.04}$ & $2.21^{+0.1}_{-0.06}$ & $0.0461^{+0.0001}_{-0.0001}$ & $-0.0^{+0.2}_{-0.2}$ & $-0.0^{+0.2}_{-0.2}$ & $0.7^{+0.2}_{-0.3}$ & $0.18^{+0.1}_{-0.04}$ & $0.65^{+0.01}_{-0.06}$\\
%127.01 & $134.03056^{+5e-05}_{-7e-05}$ & $3.5787806^{+1e-07}_{-1e-07}$ & $0.42^{+0.07}_{-0.04}$ & $2.97^{+0.09}_{-0.07}$ & $0.0996^{+0.0007}_{-0.0004}$ & $0.35^{+0.06}_{-0.07}$ & $0.23^{+0.06}_{-0.11}$ & $0.53^{+0.07}_{-0.02}$ & $0.6^{+0.1}_{-0.1}$ & $0.614^{+0.008}_{-0.01}$\\
%... &... &...  &...  &... &... &... &... &... &... &... \\
$\vdots$ &$\vdots$ &$\vdots$  &$\vdots$  &$\vdots$ &$\vdots$ &$\vdots$ &$\vdots$ &$\vdots$ &$\vdots$ &$\vdots$ \\
1793.01 & $131.7787^{+0.0005}_{-0.001}$ & $3.26176^{+1e-06}_{-1e-06}$ & $1.2^{+0.5}_{-0.9}$ & $2.8^{+1.0}_{-0.1}$ & $0.4^{+0.4}_{-0.3}$ & $-0.1^{+0.1}_{-0.1}$ & $-0.1^{+0.1}_{-0.1}$ & $0.6^{+0.3}_{-0.3}$ & $0.6^{+0.3}_{-0.4}$ & $0.8^{+0.2}_{-0.4}$\\
%KOI & KIC & $t_0$ [BKJD] & $P$ [days] & $b$ & $\log_{10}(\rho_*)$ & $R_p/R_*$ & $\sqrt{e}\cos{\omega}$ & $\sqrt{e}\sin{\omega}$ & $\alpha_r$ & $\alpha_h$ & $\alpha_{\theta}$\\
% & & &  &  & $[\mathrm{kg}/\mathrm{m}^3]$ & & & & & & \\ 
%\hline
%\endhead

\end{longtable*}
\pagebreak
\normalsize

To identify circularized KOIs, we adopt a theoretical upper limit for $\tau_{circ}$ from \cite{haswell10}, based on an upper limit for planet mass $M_{p,\ max} = 25 M_J = 0.025 M_{\odot}$ chosen to exceed the mass of any confirmed exoplanet in the exoplanets.org database: 

\small
\begin{equation}
    \tau_{circ} \leq \frac{P}{21\pi} \left(\frac{K_{dP}}{Q_P}\right)^{-1} \frac{0.025 M_{\odot}}{M_{*}} \left(\frac{a}{R_*}\right)^5 \left(\frac{R_P}{R_*}\right)^{-5}
\end{equation}
\normalsize

Here, $P$ is the orbital period of the planet, $M_*$ is the host star's mass, $R_*$ is the stellar radius, $a$ is the planet's semi-major axis, and $R_p$ is the planet's radius. $K_{dP}$ is the planet's dynamical Love number, a dimensionless parameter which expresses the ratio of the additional gravitational potential produced by tidal redistribution of the planet's mass to the gravitational potential before redistribution \citep{poulsen09,love34}. $Q_P$ is the planet's tidal quality factor, another dimensionless parameter which quantifies the efficiency of tidal dissipation in the planet \citep{ogilvie04}.

Using Kepler's third law, we may express $a/R_*$ in terms of $M_*$ and $R_*$, which are more reliably reported in the \textit{Kepler} catalog because they do not depend on transit modeling. This conversion yields

\footnotesize
\begin{equation}
    \tau_{circ} \leq \frac{P}{21\pi} \left(\frac{K_{dP}}{Q_P}\right)^{-1} \frac{0.025 M_{\odot}}{M_{*}}  \left(\frac{R_P}{R_*}\right)^{-5}\left(\frac{P^2 G M_*}{4\pi^2R_*^3}\right)^{5/3}
\end{equation}
\normalsize

We apply a linear interpolation to Solar System values to obtain the following equation for $\frac{K_{dP}}{Q_P}$ \citep{teachey17}: 

\begin{equation}
    \frac{K_{dP}}{Q_P} = 10^{-2.90 - 20.33 \frac{R_P}{R_*} \frac{R_*}{R_{\odot}}}
\end{equation}

To assemble our tidal target list, we select every KOI with $\tau_{circ}$ less than $10^8$ years according to these equations. There are 19 such KOIs; the maximum orbital period of these is 1.6 days (KOI-809.01). Of these,  we remove five from the target list: KOI-203.01, again due to its stroboscopic starspot activity \citep{desert11}; KOI-1546.01, for detected transit timing variations \citep{holczer16}; KOI-3156.01, an identified hierarchical quintuple star system \citep{shibahashi12, rappaport16}; KOI-5804.01, a highly active star where NEA-identified ``transits" correspond to alternating minima in the stellar light curve; and KOI-6534.01, which has no visible transits in its light curve at the NEA-determined transit epoch and period.

Finally, we remove 1 tidal target because because its MCMC analysis was prohibitively slow (see Section~\ref{subsec:modeling}) as a result of its unusually high number of data points ($\sim 10^5$, compared to $\sim 10^3-10^4$ for successfully analyzed tidal targets). The remaining 13 tidal targets are listed in Table~\ref{table:TC}. 

%\pagebreak
%\begin{center}
\tiny
\tabcolsep=0.015cm
%\begin{adjustbox}{width=1.2\textwidth,center}
%\begin{longtable*}[tp]{lllllllllll}
\begin{longtable*}[tp]{@{\extracolsep{\fill}}*{11}{l}@{}}

\caption{Tidally circularized (``tidal") targets, with measured transit parameters, stellar densities, and limb darkening coefficients. A machine-readable version of this table with more significant figures and the complete target list is available online. \label{table:TC}}\\

KOI & $t_0$ [BKJD] & $P$ [days] & $b$ & $\log_{10}(\rho_*$) & $R_p/R_*$ & $\sqrt{e}\cos{\omega}$ & $\sqrt{e}\sin{\omega}$ & $\alpha_r$ & $\alpha_h$ & $\alpha_{\theta}$\\
 & &  &  & $[\mathrm{kg}/\mathrm{m}^3]$ & & & & & & \\
\hline
\endhead
809.01 & $170.64828^{+7e-05}_{-0.00186}$ & $1.5947455^{+1e-07}_{-1e-07}$ & $0.67^{+0.04}_{-0.02}$ & $3.1^{+0.03}_{-0.02}$ & $0.13^{+0.002}_{-0.002}$ & $-0.02^{+0.07}_{-0.56}$ & $-0.03^{+0.08}_{-0.15}$ & $0.3^{+0.4}_{-0.2}$ & $0.5^{+0.3}_{-0.2}$ & $0.22^{+0.31}_{-0.08}$\\
1064.01 & $133.49^{+0.21}_{-0.09}$ & $1.187^{+0.005}_{-0.005}$ & $0.8^{+0.3}_{-0.1}$ & $1.97^{+0.55}_{-0.08}$ & $0.5^{+0.2}_{-0.3}$ & $0.0^{+0.3}_{-0.2}$ & $-0.1^{+0.2}_{-0.2}$ & $0.7^{+0.3}_{-0.2}$ & $0.6^{+0.3}_{-0.3}$ & $0.25^{+0.34}_{-0.09}$\\
1075.01 & $133.2764^{+0.0004}_{-0.0007}$ & $1.3437661^{+6e-08}_{-7e-08}$ & $1.91^{+0.02}_{-0.04}$ & $2.437^{+0.007}_{-0.011}$ & $0.98^{+0.01}_{-0.04}$ & $-0.01^{+0.03}_{-0.04}$ & $-0.1^{+0.02}_{-0.02}$ & $0.8^{+0.2}_{-0.4}$ & $4e-05^{+0.00094}_{-4e-05}$ & $0.07^{+0.86}_{-0.05}$\\
%1658.01 & $133.54889^{+0.00016}_{-6e-05}$ & $1.5449288^{+1e-07}_{-3e-07}$ & $0.495^{+0.053}_{-0.001}$ & $3.673^{+0.002}_{-0.819}$ & $0.0687^{+0.002}_{-0.0002}$ & $-0.08^{+0.02}_{-0.04}$ & $0.424^{+0.0007}_{-0.1442}$ & $8e-05^{+0.35166}_{-8e-05}$ & $0.513^{+0.003}_{-0.144}$ & $0.57^{+0.03}_{-0.09}$\\
%2925.01 & $131.721^{+0.006}_{-0.003}$ & $0.716532^{+2e-06}_{-3e-06}$ & $1.0^{+0.4}_{-0.7}$ & $3.7^{+0.7}_{-1.5}$ & $0.017^{+0.396}_{-0.009}$ & $0.01^{+0.05}_{-0.06}$ & $-0.02^{+0.07}_{-0.07}$ & $0.5^{+0.3}_{-0.3}$ & $0.5^{+0.3}_{-0.3}$ & $0.6^{+0.3}_{-0.4}$\\
%3913.01 & $131.621^{+0.0007}_{-0.0006}$ & $0.5828955^{+3e-07}_{-3e-07}$ & $1.1^{+0.6}_{-0.8}$ & $2.8^{+1.1}_{-0.4}$ & $0.1^{+0.6}_{-0.1}$ & $0.0^{+0.06}_{-0.06}$ & $0.0^{+0.06}_{-0.06}$ & $0.6^{+0.3}_{-0.4}$ & $0.6^{+0.3}_{-0.3}$ & $0.7^{+0.2}_{-0.5}$\\
%4351.01 & $131.9546^{+0.0009}_{-0.0008}$ & $0.6443225^{+2e-07}_{-2e-07}$ & $1.4^{+0.4}_{-0.4}$ & $2.8^{+0.3}_{-0.3}$ & $0.5^{+0.4}_{-0.4}$ & $0.05^{+0.11}_{-0.08}$ & $-0.04^{+0.1}_{-0.09}$ & $0.7^{+0.3}_{-0.5}$ & $0.9^{+0.1}_{-0.3}$ & $0.6^{+0.3}_{-0.3}$\\

%... &... &...  &...  &... &... &... &... &... &... &... \\
$\vdots$ &$\vdots$ &$\vdots$  &$\vdots$  &$\vdots$ &$\vdots$ &$\vdots$ &$\vdots$ &$\vdots$ &$\vdots$ &$\vdots$ \\
7449.01 & $132.62^{+0.03}_{-0.14}$ & $1.3245^{+0.0029}_{-0.0006}$ & $1.2^{+0.8}_{-0.3}$ & $1.4^{+0.5}_{-0.2}$ & $0.4^{+0.6}_{-0.3}$ & $0.03^{+0.15}_{-0.08}$ & $-0.0^{+0.3}_{-0.2}$ & $0.3^{+0.6}_{-0.3}$ & $0.71^{+0.09}_{-0.44}$ & $0.6^{+0.1}_{-0.2}$\\

%KOI & KIC & $t_0$ [BKJD] & $P$ [days] & $b$ & $\log_{10}(\rho_*)$ & $R_p/R_*$ & $\sqrt{e}\cos{\omega}$ & $\sqrt{e}\sin{\omega}$ & $\alpha_r$ & $\alpha_h$ & $\alpha_{\theta}$\\
% & & &  &  & $[\mathrm{kg}/\mathrm{m}^3]$ & & & & & & \\ 
%\hline
%\endhead

\end{longtable*}

%\end{adjustbox}
%\end{center}

%%%\pagebreak
\normalsize

\subsubsection{Compact multi-planet systems}\label{subsubsec:MP}

Finally, we consider compact multi-planet systems, which are not expected to be dynamically stable unless their constituent planets are on low-eccentricity orbits. \cite{vea15} quantify this expectation by examining 28 $Kepler$ multi-planet host stars with asteroseismic $\rho_*$ measurements. They use the discrepancy between the asteroseismic and transit-derived $\rho_*$ measurement to measure the eccentricity of each of the 74 KOIs in their sample. They find that the resulting eccentricities are well described by a Rayleigh distribution, $P(e,\sigma) = \frac{e}{\sigma^2}\exp{\frac{-e^2}{2\sigma^2}}$, with $\sigma = 0.049 \pm 0.013$.

We cannot properly use this Rayleigh distribution as a prior $e$ distribution to measure $\rho_*$ of the host stars in the \cite{vea15} sample itself, because the $\rho_*$ information contained in those KOIs' transits was used to define the prior in the first place. Rather, we must identify an independent sample of KOIs which resembles the sample of \cite{vea15}. 

To assemble this sample, we compare the distribution of period ratios of the \cite{vea15} sample to that of the remaining \textit{Kepler} multi-planet systems. For each multi-planet system, we calculate the ratio of the orbital period of each outer planet to its nearest inner neighbor. The distribution of period ratios in the \cite{vea15} sample serves as a reference distribution; we identify a sample of 1340 KOIs which is consistent with the \cite{vea15} sample by an Anderson-Darling test ($p = 0.4$; \citealt{ad52}). We further subject the two samples to a Kolmogorov-Smirnov test and find them consistent at the $p=0.15$ level \citep{k33,s48}. 

We impose a signal-to-noise cutoff on this KOI sample, discarding planetary systems where the \textit{Kepler}-reported transit model SNR $< 50$ for one or more of the KOIs. After the cutoff, 27 systems remain;  of these 27 systems, 12 exhibit TTVs \citep{holczer16} and are removed from the target list. We remove a further 2 systems, comprising 5 KOIs, because because their MCMC analysis was prohibitively slow (see Section~\ref{subsec:modeling}) as a result of their unusually high number of data points ($\sim 10^5-10^6$, compared to $\sim 10^3-10^5$ for successfully analyzed multi-planet targets), and 4 further systems for having very few remaining posterior samples after we perform some quality checks (see Section~\ref{sec:results} for details). The remaining 9 systems, comprising 18 KOIs, are listed in Table~\ref{table:MP}.

%\pagebreak
%\begin{center}
\tiny
\tabcolsep=0.015cm
%\begin{adjustbox}{width=1.2\textwidth,center}
%\begin{longtable*}[tp]{lllllllllll}
\begin{longtable*}[tp]{@{\extracolsep{\fill}}*{11}{l}@{}}

\caption{Compact multi-planet systems (``multis") selected to be statistically consistent with the sample of \cite{vea15}, with measured transit parameters, stellar densities, and limb darkening coefficients. A machine-readable version of this table with more significant figures and the complete target list is available online.  \label{table:MP}}\\

KOI & $t_0$ [BKJD] & $P$ [days] & $b$ & $\log_{10}(\rho_*$) & $R_p/R_*$ & $\sqrt{e}\cos{\omega}$ & $\sqrt{e}\sin{\omega}$ & $\alpha_r$ & $\alpha_h$ & $\alpha_{\theta}$\\
 & &  &  & $[\mathrm{kg}/\mathrm{m}^3]$ & & & & & & \\
\hline
\endhead
124.01 & $137.136^{+0.004}_{-0.011}$ & $12.69085^{+0.00025}_{-5e-05}$ & $0.68^{+0.31}_{-0.05}$ & $2.8^{+0.3}_{-1.5}$ & $0.0154^{+0.0066}_{-0.0005}$ & $0.25^{+0.04}_{-0.01}$ & $0.1^{+0.06}_{-0.03}$ & $0.87^{+0.02}_{-0.02}$ & $0.58^{+0.04}_{-0.03}$ & $0.3^{+0.01}_{-0.17}$\\
124.02 & $142.833^{+0.008}_{-0.004}$ & $31.71962^{+0.00025}_{-5e-05}$ & $0.83^{+0.09}_{-0.03}$ &   & $0.019^{+0.0886}_{-0.0004}$ & $0.14^{+0.1}_{-0.06}$ & $-0.05^{+0.08}_{-0.29}$ &   &   &  \\
153.01 & $139.7138^{+0.0005}_{-0.0011}$ & $8.925083^{+2e-06}_{-9e-06}$ & $0.98^{+0.04}_{-0.01}$ & $1.95^{+0.05}_{-0.02}$ & $0.11^{+0.02}_{-0.03}$ & $0.03^{+0.02}_{-0.02}$ & $-0.246^{+0.009}_{-0.009}$ & $0.65^{+0.06}_{-0.03}$ & $0.51^{+0.09}_{-0.04}$ & $0.991^{+0.007}_{-0.011}$\\
153.02 & $128.72^{+0.03}_{-0.03}$ & $4.754008^{+2e-06}_{-9e-06}$ & $1.06^{+0.04}_{-0.06}$ &   & $0.08^{+0.04}_{-0.05}$ & $-0.16^{+0.04}_{-0.03}$ & $-0.513^{+0.006}_{-0.009}$ &   &   &  \\
678.01 & $172.61^{+0.3}_{-0.01}$ & $6.0634^{+0.0001}_{-0.0201}$ & $0.02^{+0.84}_{-0.02}$ & $0.7^{+1.9}_{-0.2}$ & $0.023^{+0.024}_{-0.001}$ & $0.14^{+0.02}_{-0.01}$ & $0.4^{+0.2}_{-0.3}$ & $0.18^{+0.13}_{-0.02}$ & $0.76^{+0.01}_{-0.2}$ & $0.64^{+0.08}_{-0.44}$\\
678.02 & $132.04^{+0.41}_{-0.08}$ & $4.1615^{+0.0001}_{-0.0201}$ & $1.0^{+0.01}_{-0.23}$ &   & $0.02^{+0.03}_{-0.02}$ & $-0.37^{+0.03}_{-0.04}$ & $-0.24^{+0.02}_{-0.02}$ &   &   &  \\

$\vdots$ &$\vdots$ &$\vdots$  &$\vdots$  &$\vdots$ &$\vdots$ &$\vdots$ &$\vdots$ &$\vdots$ &$\vdots$ &$\vdots$ \\
2687.01 & $131.96^{+0.07}_{-0.22}$ & $1.717^{+0.016}_{-0.003}$ & $1.0^{+0.5}_{-0.1}$ & $2.9^{+1.2}_{-2.0}$ & $0.1^{+0.4}_{-0.1}$ & $0.0^{+0.1}_{-0.1}$ & $-0.34^{+0.18}_{-0.08}$ & $0.42^{+0.3}_{-0.08}$ & $0.6^{+0.1}_{-0.3}$ & $0.7^{+0.2}_{-0.4}$\\
2687.02 & $136.2^{+0.4}_{-0.4}$ & $8.167^{+0.016}_{-0.003}$ & $1.0^{+0.4}_{-0.3}$ &   & $0.3^{+0.2}_{-0.3}$ & $-0.2^{+0.3}_{-0.1}$ & $-0.2^{+0.4}_{-0.4}$ &   &   &  \\

%KOI & KIC & $t_0$ [BKJD] & $P$ [days] & $b$ & $\log_{10}(\rho_*)$ & $R_p/R_*$ & $\sqrt{e}\cos{\omega}$ & $\sqrt{e}\sin{\omega}$ & $\alpha_r$ & $\alpha_h$ & $\alpha_{\theta}$\\
% & & &  &  & $[\mathrm{kg}/\mathrm{m}^3]$ & & & & & & \\ 
%\hline
%\endhead

\end{longtable*}

%\end{adjustbox}
%\end{center}

%%%\pagebreak
\normalsize

\subsection{Detrending}\label{subsec:detrending}
Here, we describe our procedure for detrending the \textit{Kepler} light curves of our 75 target KOIs (orbiting 66 target stars) in preparation for transit modeling. The trends in question are due to stellar activity or instrumental effects and are superimposed on the planetary transits in the light curve.

\subsubsection{Outlier removal}\label{subsubsec:outliers}
We begin by splitting each target KOI's full \textit{Kepler} simple aperture photometry light curve into individual transits, each bookended by sufficient out-of-transit observation time to capture out-of-transit trends in the light curve. For targets with available short-cadence observations (58.86 seconds per exposure), we apply the below procedure to both short- and long-cadence data; otherwise we use long-cadence data (29.4 minutes per exposure).

To slice the light curve, we use the NEA-reported transit ephemeris $t_0$, orbital period $P$, and transit duration $T_{14}$ \citep{akeson15}. We divide each light curve into segments centered at $t_0$ plus successive integer values of $P$. For each segment, we keep out-of-transit data spanning an interval $t_{1/2} + t_{OOT}$ on either side of $t_0$, where we define $t_{1/2}$ as slightly more than half a transit duration, and $t_{OOT}$ as an ``out-of-transit window:"

\begin{equation}
    t_{1/2} = 1.1 \left(\frac{T_{14}}{2} + t_{LC}\right)
\end{equation}

\begin{equation}
    t_{OOT} = \sqrt{10}\,t_{1/2},
\end{equation}

or roughly $3\,t_{1/2}$. Here, $t_{LC}$ is the integration time of a long-cadence \textit{Kepler} exposure, equal to 29.4 minutes. We discard data points where $|t - t_0| > (t_{1/2} + t_{OOT})$.

Once each KOI's light curve is divided into individual transit segments, we remove flux outliers and discard transit segments with insufficient data. To remove outlying data points within each transit segment, we perform a moving median smoothing of the out-of-transit (i.e. $|t - t_0| > t_{1/2}$) flux data points, with a kernel size of 21 data points. We then reject any data points more than $3\sigma$ away from the moving median-smoothed light curve, where $\sigma$ is defined as the \textit{Kepler}-reported uncertainty of each flux measurement. A small number of transit segments also exhibit clear outliers within $t_{1/2}$ of a segment midpoint, identifiable as data points with anomalously high flux. We remove any data point that lies more than $3\sigma$ \textit{above} the within-transit light curve.

Finally, after removing individual outlying data points, we reject any full transit segment where one of the following conditions is met:
\begin{enumerate}
\item There are fewer than 3 out-of-transit data points on one side of $t_0$;
\item There are more than 3 out-of-transit data points, but they span a very short time interval (i.e., less than $2\,t_{LC}$); or
\item The out-of-transit data points immediately adjacent to the transit are missing. Such missing data could lead to poor constraints on the transit depth or duration.
\end{enumerate}

After outlier removal, each target KOI's light curve is reduced to a series of individual transit segment light curves. Each transit observed at \textit{Kepler's} long cadence contains $\sim 30$ data points, and each transit observed in short cadence contains $\sim 750$ data points.

\subsubsection{Evaluation of out-of-transit trends}
The transit segments isolated by the above procedure are individually afflicted by out-of-transit trends due to stellar activity and instrumental variation. To fit a precise transit model to each KOI, we must first account for these trends (e.g. \citealt{aigrain16,luger16}). One approach to detrending would be to add additional free parameters to our transit model to describe each transit's trends individually; fit them all; and then marginalize over them to recover the physically interesting parameters describing the planet-star system. However, this would add prohibitive computational cost, all for the sake of nuisance parameters.

Instead, we elect to detrend each transit segment using linear least squares regression (see e.g. \citealt{kundurthy11}). We assume that the out-of-transit trend for each transit segment is well-fit by a low-order polynomial of predetermined order, then divide out the best-fitting polynomial trend at each MCMC step before calculating the likelihood of the transit model parameters. Polynomial detrending is a common approach to analyzing \textit{Kepler} data (see e.g. \citealt{oleary14, fabrycky12, orosz12, lissauer11}).

To choose the appropriate polynomial order for each transit segment, we use the Bayesian Information Criterion (BIC), a model selection statistic which balances goodness-of-fit against the number of free parameters in the model, i.e. the polynomial order:

\begin{equation}
    BIC = \chi^2 + k \ln{n}
\end{equation}

Here, $k$ is the number of free parameters in the model, $n$ is the number of data points, and $\chi^2$ is the squared error of the model, scaled by the measurement uncertainties.

%:
%\begin{equation}
%    \chi^2 = \sum_{i=1}^{n} \frac{(y_i - f(x_i))^2}{\sigma_{y_i}^2}
%\end{equation}

%Here, $y_i$ are the observed flux values, $f(x_i)$ are the polynomial model-predicted flux values calculated for the time coordinates $x_i$, and $\sigma_{y_i}$ are the uncertainties on the observed fluxes. 

For each transit segment, with the in-transit data masked, we test polynomials of orders ranging from 0 to 3 and select the polynomial model with the lowest BIC. At each MCMC step (see section~\ref{subsec:modeling}, below), before evaluating the likelihood of the transit model calculated from the sampled parameters, we (i) calculate, analytically, the best-fitting polynomial of this pre-selected order for each transit segment using linear least squares regression and (ii) impose this best-fitting polynomial trend upon the transit model. We are then evaluating the likelihood of the transit model given the data, both subject to the same out-of-transit trends.

\subsection{Transit modeling}\label{subsec:modeling}

With this polynomial-fitting procedure in place, we
explore the parameter space of our transit model to identify the region that describes each planet's transit light curve best. For the occultation and tidal targets, this space is ten-dimensional. For the multi-planet targets, it is $(4 + 6N)$-dimensional, where $N$ is the number of planets in the system; 4 parameters describe the star and are the same for every KOI in the system ($\rho_*$ and the three LDCs, $\alpha_r$, $\alpha_h$, and $\alpha_\theta$), and 6 describe each KOI (the transit epoch $t_0$, the period $P$, the impact parameter $b$, the ratio-of-radii $R_p/R_*$, and the reparametrized eccentricity and argument of periastron, $\sqrt{e}\cos{\omega}$ and $\sqrt{e}\sin{\omega}$).

We evaluate the likelihood of any given set of $4 + 6N$ transit parameters by using the transit modeling package \texttt{BATMAN} \citep{kreidberg15} to calculate a light curve directly from the parameters. To this calculated light curve, we calculate and apply the best-fitting out-of-transit polynomial trend of pre-determined order (see Section~\ref{subsec:detrending}) for each observed transit of the target KOI to enable a direct comparison of the model to the data. We then calculate the likelihood of the data given the transit model parameters. We adopt a Gaussian likelihood function. 

We explore the $4+6N$-dimensional parameter space of the transit model with the affine-invariant MCMC ensemble sampler package \texttt{emcee}. \texttt{emcee} initializes an ensemble of MCMC walkers in this parameter space and calculates the posterior probability of the sampled set of transit parameters at every step in their random walk, given a choice of prior distributions and our Gaussian likelihood function. 

%By Bayes' theorem, this posterior probability of the parameters $\boldsymbol{\Theta}$ is proportional to the prior probability of these parameters times the likelihood of the data $\mathcal{D}$ given the parameters:

%\begin{equation}
%\overbrace{\mathrm{Pr}(\boldsymbol{\Theta}|\mathcal{D})}^{\mathrm{posterior}} \propto \overbrace{\mathrm{Pr}(\mathcal{D}|\boldsymbol{\Theta})}^{\mathrm{likelihood}} \overbrace{\mathrm{Pr}(\boldsymbol{\Theta})}^{\mathrm{prior}}
%\end{equation}

%We adopt a Gaussian likelihood function, and at each MCMC step, we calculate this likelihood by generating a transit light curve for each planet from its 10 sampled transit parameters using the transit-modeling package \texttt{BATMAN} \citep{kreidberg15} and comparing it to the data.

We adopt the following priors for the transit parameters:

\begin{enumerate}
\item Intrinsic priors:
    \begin{enumerate}
    \item $t_0$: A uniform prior from $t_{0,\,reported} - 0.5\,\textrm{days}$ to $t_{0,\,reported} + 0.5\,\textrm{days}$ , where $t_{0,\,reported}$ is the transit epoch reported in the \textit{Kepler} catalog.
    \item $P$: A uniform prior from $0.9\,P_{reported}$ to $1.1\,P_{reported}$, where $P_{reported}$ is the orbital period reported in the \textit{Kepler} catalog.
    \item $b$: A uniform prior from 0 to 2, allowing for grazing transits.
    \item $\rho_*$, reparametrized as $\log_{10}(\rho_* \mathrm{[kg/m^3]})$: A uniform prior in $\log_{10}(\rho_* \mathrm{[kg/m^3]})$ from 0 to 6.
    \item $R_p/R_*$: A uniform prior from 0 to 1.
    \item $e$ and $\omega$: Uniform priors from -1 to 1 in $\sqrt{e}\cos{\omega}$ and $\sqrt{e}\sin{\omega}$, with additional uniform priors restricting $e$ to the range $(0,1)$ and $\omega$ to the range $(-\pi, \pi)$.
    \item Nonlinear limb-darkening coefficients $\alpha_r, \alpha_h, \alpha_{\theta}$: Uniform priors from 0 to 1 \citep{kipping16ldc}.
    \item A prior insisting that $b$ be less than $(a/R_*)$, calculated from Kepler's third law, in order to prevent unphysical inclinations.
    \item A prior insisting that $b$ be less than $(1 + R_p/R_*)$, in order to prevent unphysical transit durations.
    \end{enumerate}
\item Target selection-motivated $e$ and $\omega$ priors:
    \begin{enumerate}
    \item For the occultation targets, which have secondary eclipse-measured constraints on $e$ and $\omega$, we adopt Gaussian priors in $e\cos{\omega}$ and $e\sin{\omega}$, where the means are given by the measured values of $e\cos{\omega}$ and $e\sin{\omega}$ from \cite{shabram16} or \cite{clm12} and the standard deviations by their measurement uncertainties.
    \item For the tidal targets, we adopt an exponential prior on $e$, with scale parameter $\lambda = 0.00796$, consistent with the findings of \cite{wang11}.
    \item For the multi-planet targets, we adopt a Rayleigh prior in $e$, with scale parameter $\sigma = 0.049$, consistent with the findings of \cite{vea15}.
    \end{enumerate}
\end{enumerate}

With \texttt{emcee}, we initialize 100 MCMC walkers per KOI and run them for $10^5$ steps each, generating $10^7$ posterior samples per KOI. We initialize the walkers in the $P$-dimension by drawing from a Gaussian distribution centered at the \textit{Kepler} catalog-reported $P$, with standard deviation 0.01. We initialize the walkers in the other 9 dimensions of parameter space by sampling randomly in a 9-dimensional box spanning the range in each parameter that is allowed by its intrinsic prior.

%At each step, we evaluate the transit light curve from the sampled transit parameters with \texttt{BATMAN}, apply the best-fitting out-of-transit polynomial trend (see Section~\ref{subsec:detrending}) to this light curve, and calculate the likelihood of the data given those sampled parameters by comparing the resulting light curve to the data.

We discard the first 20,000 steps per walker chain as ``burn-in," based on a conservative by-eye judgment of when the walkers ``forget" their initial conditions and begin to explore the parameter space freely. We also discard walker chains which fail to converge to the same value of $P$ as the majority of the ensemble of walkers. More specifically, we calculate the median and median absolute deviation (MAD) of $P$ over all the walker chains. We use  $\hat{\sigma} = 1.4826\times \textrm{MAD}$ as an estimator for the standard deviation of the $P$ distribution and discard any chain whose median $P$ differs from the overall median by more than $5 \hat{\sigma}$ \citep{rousseeuw93}.

Finally, for KOIs which are confirmed per their NEA disposition (two-thirds of our targets, or 50 KOIs), we discard all posterior samples with $R_p/R_* > 0.15$ on the grounds that they represent unphysically large planets. We note that all of our target planets which are dispositioned as ``confirmed" in the NEA are validated by \cite{morton16}.

\section{Results}\label{sec:results}
We obtain successful transit fits (i.e., MCMC convergence) for 66 target stars (hosting 75 individual KOIs). For four multi-planet targets (those hosting KOIs 156.01, 156.02, and 156.03; 723.01, 723.02, and 723.03; KOIs 1805.01, 1805.02, and 1805.03; and KOIs 1824.01 and 1824.02), less than a few hundred posterior samples for each system remain after we discard chains that fail to converge in $P$ and samples with unphysically large $R_p/R_*$; we count these as failed fits. The best-fit transit parameters for the 66 successes are listed in Tables \ref{table:SE}-\ref{table:MP}; we present the median of the posterior distribution for each parameter, with uncertainty bounds describing the 16th and 84th percentiles.

Of order $10^6-10^7$ samples from the posterior distributions of the transit parameters remain for each KOI after we discard the burn-in phase of the MCMC chain, as well as chains which fail to converge in $P$ and samples with unphysically large $R_p/R_*$. The files containing all of the posterior samples are prohibitively large to be made available for online download, so we  downsample the posteriors by a factor of $10^2$ and publish the resulting $10^4-10^5$ posterior samples for each KOI at  \url{https://doi.org/10.5281/zenodo.1028515}. 

\begin{figure*}[!ht]
\begin{center}
\includegraphics[width=\textwidth]{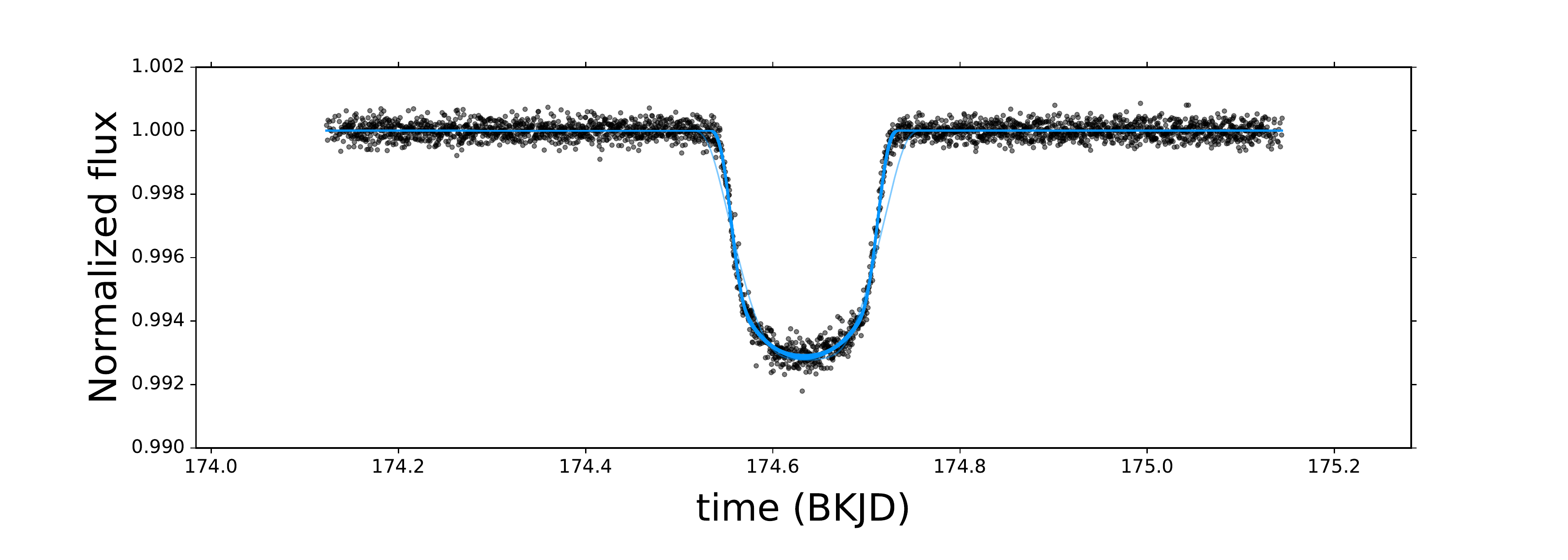}
\caption{An example transit fit, for occultation target KOI-929.01, a confirmed planet per its NEA disposition. The black points are the 183 transits observed for this KOI, detrended and stacked; it is not observed in short cadence, so all of these data points are long-cadence observations. The blue lines are light curve models computed by \texttt{BATMAN} \citep{kreidberg15} from 500 random draws from our 10-dimensional transit parameter posterior distributions. This KOI has an orbital period of $6.491683\pm0.000002$ days, and it orbits a star of \textit{Kepler}-band magnitude 15.649.}
\label{fig:K00929batman}
\end{center}
\end{figure*}

As an example, in Figure~\ref{fig:K00929batman}, we present a well-converged transit fit, for occultation target KOI-929.01 (a confirmed planet, per the NEA). We plot the corresponding posterior distributions for the 10 fitted transit parameters in Figure~\ref{fig:K00929post}. From each posterior sample, we calculate nine other parameters describing the system (the transit duration $T_{14}$, the flat-bottomed transit duration $T_{23}$, the normalized semi-major axis $a/R_*$, the inclination $i$, the eccentricity $e$, the argument of periastron $\omega$, and the three traditional nonlinear limb-darkening coefficients $c_2$, $c_3$, and $c_4$). We plot the distributions of these derived parameters in Figure~\ref{fig:K00929derivedpost}.

KOI-929.01's posterior distributions typify our broader results: $t_0$ and $P$ are by far the most precisely constrained parameters, $\rho_*$ is constrained to within $5\%$ of its median value, and $\sqrt{e}\cos{\omega}$ and $\sqrt{e}\sin{\omega}$ are centered at zero, in agreement with the prior constraints on this planet's orbit from secondary eclipse observations. When we derive the distributions of $e$ and $\omega$ themselves (see Figure~\ref{fig:K00929derivedpost}), we find that $e$ is strongly peaked at $e=0$, and $\omega$ is very poorly constrained, which is sensible for a nearly-circular orbit.  Also typical are the constraints on the three limb-darkening coefficients: $\alpha_{\theta}$ is constrained to within $\sim 15 \%$ of its median value, and $\alpha_r$ and $\alpha_h$ only to within $\sim 60 \%$.

\subsection{Covariances}\label{subsec:covariances}

In this section, we investigate covariances between the transit parameters, which indicate degeneracies in the transit model. In other words, if independently adjusting two or more of the parameters can create the same effect in the shape of the model light curve, these parameters will correlate with each other, or co-vary.

\begin{figure*}[!ht]
\begin{center}
\includegraphics[width=\textwidth]{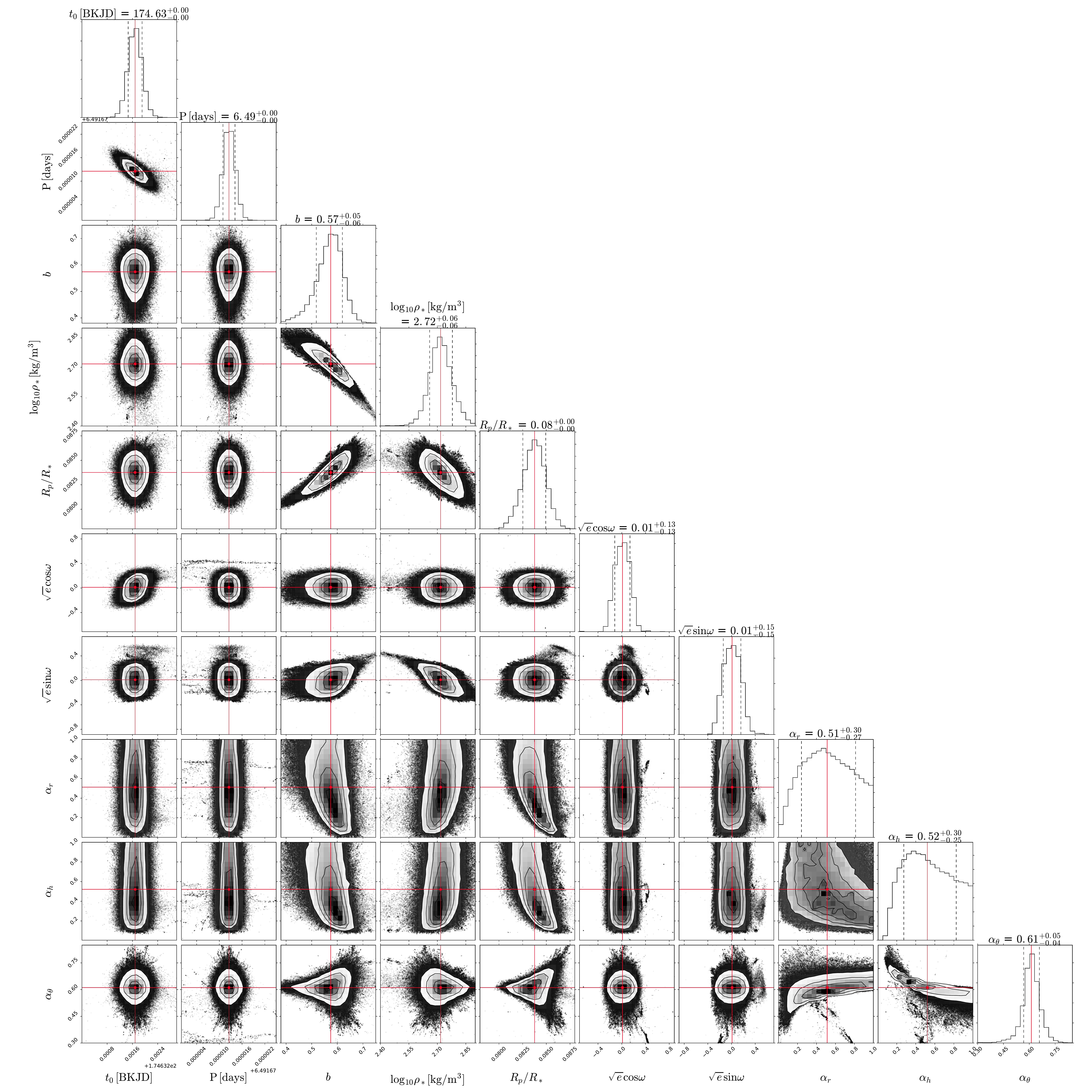}
\caption{Posterior distributions for the ten fitted transit parameters of occultation target KOI-929.01, a 
confirmed planet per the NEA. The red lines mark the median of each distribution; the black dotted lines mark the 16th and 84th percentiles.}
\label{fig:K00929post}
\end{center}
\end{figure*}

A well-known effect in transit modeling (see e.g. \citealt{carter08}) is the covariance between stellar density $\rho_*$, impact parameter $b$, and ratio-of-radii $R_p/R_*$, which results from the mixed influence of these three parameters on the transit duration. For example, a larger $R_p/R_*$, a smaller $b$, and a lower $\rho_*$ all lead to a longer transit duration. This covariance manifests itself in the posterior distributions of several of our less-well-constrained targets, especially those for which no short-cadence observations are available and those which are not confirmed per the NEA (for which we cannot discard posterior samples with $R_p/R_* > 0.15$). For such targets, our posterior plots show an elongated positive correlation between the $b$ and $R_p/R_*$ distributions, as well as a tail of low $b$ values which are negatively correlated with $\log_{10}\rho_*$. Figure~\ref{fig:K00929post}, although it is a confirmed planet per the NEA, exhibits these trends.

The ultimate overall effect of this covariance is a distinct bimodality in each of the $b$, $\log_{10}\rho_*$, and $R_p/R_*$ posterior distributions. For physical intuition, this bimodality signifies that two transit models are likely given the observations: one in which a relatively small planet undergoes a non-grazing transit across a compact star, and one in which a relatively large planet undergoes a grazing transit across a large, low-density star.

\textit{A-priori}, the high-$b$, low-$\log_{10}\rho_*$, high-$R_p/R_*$ peak is physically implausible, on the grounds that we are much more likely to observe a small planet transiting across the midpoint of its host star than we are to observe an enormous planet transiting across the limb \citep{kippingsandford16}. For independently confirmed KOIs, we exclude all posterior samples with $R_p/R_* > 0.15$ on these physical grounds, because anything larger than this approximate limit would be an eclipsing binary, not a transiting planet.  We cannot, however, exclude the large $R_p/R_*$ samples for KOIs which are not formally confirmed per the NEA, in the case that they turn out to be eclipsing binaries. 

A strong intrinsic covariant prior on $R_p/R_*$ and $\log_{10}\rho_*$, i.e., a way to formally encode our skepticism of grazing, large-planet fits, would address this problem of bimodality, as discussed in \citealt{kippingsandford16}. However, the exoplanet population data are not yet robust enough to define such a prior.

Also evident in the posterior distributions of the transit parameters is a covariance between the limb darkening coefficients $\alpha_r$ and $\alpha_h$. For the vast majority of our target stars, these parameters are not tightly constrained---the posteriors displayed in Figure~\ref{fig:K00929post} are typical. Although the peaks of the distributions of $\alpha_r$ and $\alpha_h$ are broad, however, there is a clear negative correlation between the two, with high $\alpha_r$ corresponding to low $\alpha_h$ and vice versa. This covariance explains the trends we discuss in Section~\ref{subsubsec:AS}, where we compare our observed $\alpha$s to theoretical predictions from stellar atmosphere modeling.

\begin{figure*}[!ht]
\begin{center}
\includegraphics[width=\textwidth]{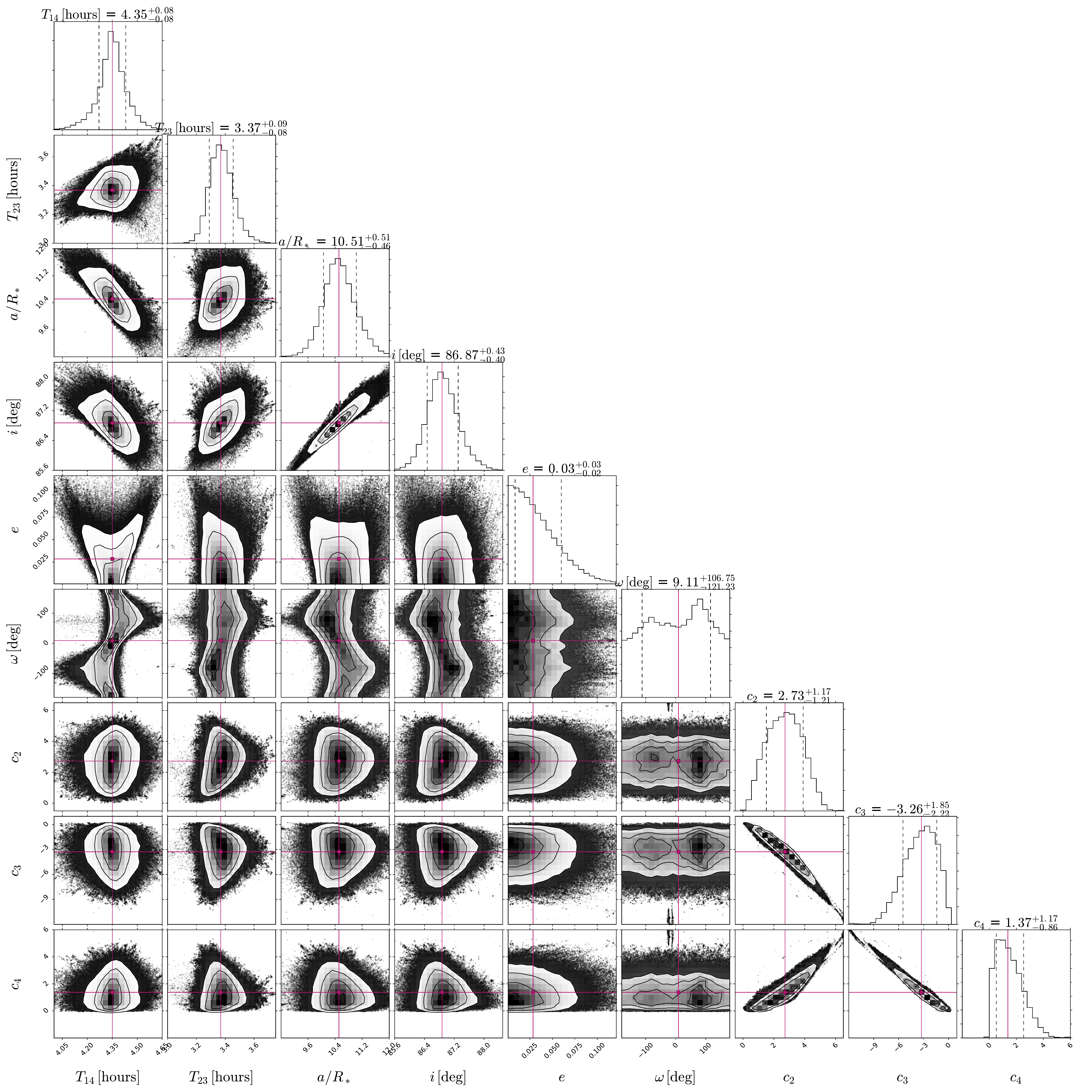}
\caption{Distributions for nine derived parameters of occultation target KOI-929.01, a confirmed planet per the NEA. These are parameters which we did not fit for directly but can compute from the posterior samples plotted in Figure~\ref{fig:K00929post}. Here, $c_2$, $c_3$, and $c_4$ are the traditional coefficients of a modified nonlinear limb-darkening law, computed from our reparametrized $\alpha$s. The purple lines mark the median of each distribution; the black dotted lines mark the 16th and 84th percentiles.}
\label{fig:K00929derivedpost}
\end{center}
\end{figure*}

Finally, we note a strong covariance between $c_2$, $c_3$, and $c_4$, evident in the rightmost panels of Figure~\ref{fig:K00929derivedpost}. This covariance, which motivated the transformation to $\alpha$-space originally \citep{kipping16ldc}, exists because only a relatively small region of the three-dimensional $c$-space describes physically realistic limb-darkening behavior. 

\subsection{Stellar densities}\label{subsec:rhostar_results}

\begin{figure*}[!pht]
\begin{center}
\includegraphics[width=0.66\textwidth]{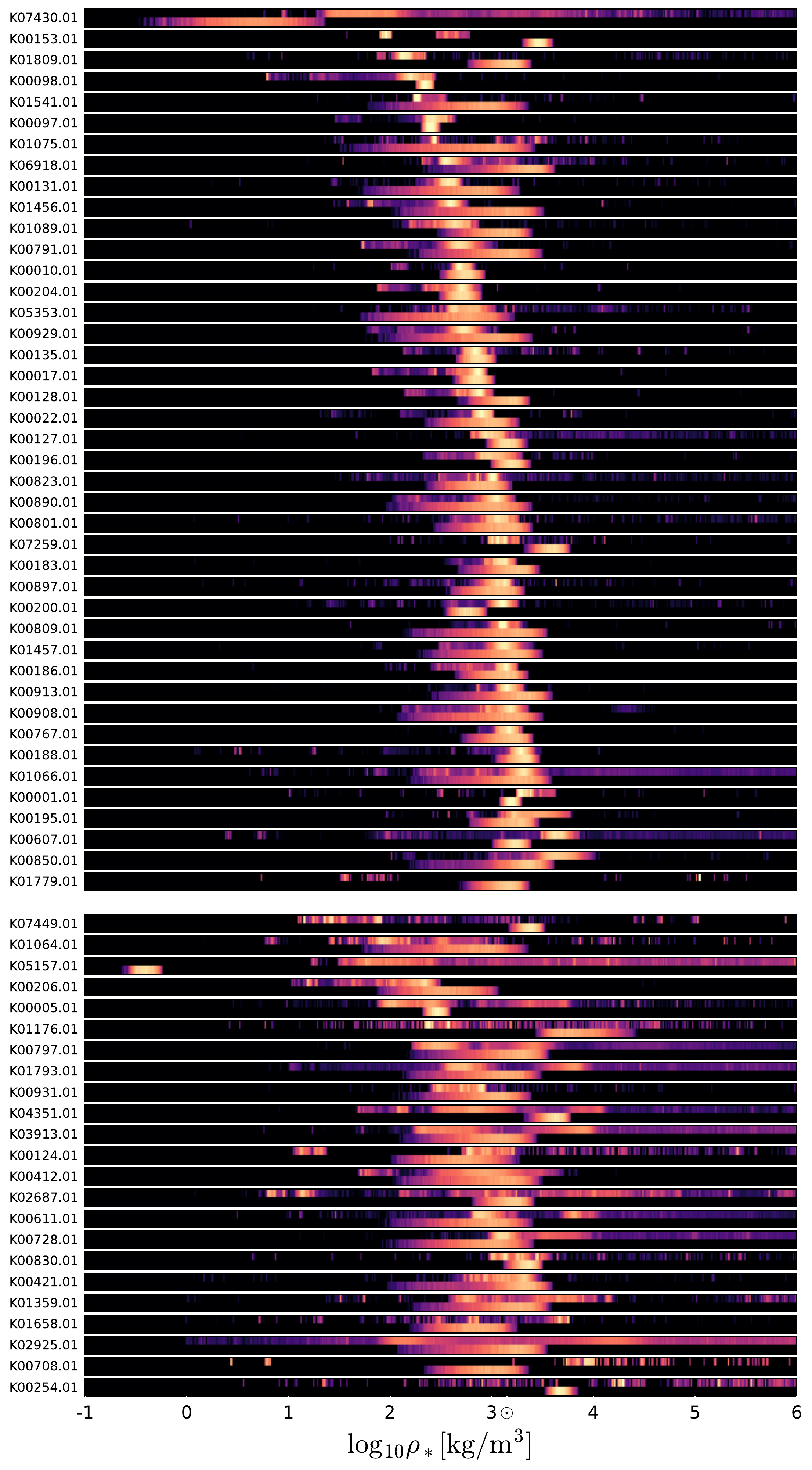}
\caption{A comparison of our $\rho_*$ posteriors to previously published constraints. Top block: KOIs for which we achieve $\leq 5\%$ fractional uncertainty on $\log_{10}\rho_*$ ($62\%$ of targets); bottom block: KOIs for which we do not. For each KOI, the upper row shows the posterior distribution of $\log_{10}\rho_*$ derived in this work, and the lower row shows the \textit{Kepler} Data Release 25 constraint \citep{mathur17} on $\log_{10}\rho_*$ derived from Dartmouth Stellar Evolution Database isochrone modeling. Within each block, the KOIs are sorted from top to bottom in order of increasing median $\log_{10}\rho_*$ from our results.}
\label{fig:rhostar_precision}
\end{center}
\end{figure*}

%tktk relabel this figure for the multis, to make it clear that the row corresponds to every KOI in the system?
The ensemble results of our stellar density measurements are presented in Figure~\ref{fig:rhostar_precision}. In this figure, for each target star, we compare the posterior distributions of $\log_{10}\rho_*$ derived from our transit fitting with the constraints on $\log_{10}\rho_*$ from the \textit{Kepler} Data Release 25 (DR25) Stellar Properties Catalog by \cite{mathur17}. The posteriors from the \cite{mathur17} catalog are derived by performing Dartmouth Stellar Evolution Database isochrone modeling on input values of $T_{\textrm{eff}}$, $\log{g}$, and [Fe/H] obtained from earlier studies relying on a variety of experimental methods, including spectroscopy, flicker, asteroseismology, and previous transit modeling.

Our transit modeling-derived $\log_{10}\rho_*$ is in $1\sigma$ or better agreement with the isochrone modeling-derived DR25 constraint for $55\%$ of our target stars, and in $3\sigma$ agreement for $95\%$. The only three target stars for which we disagree with the DR25 stellar density constraint at the $3\sigma$ level are tidal targets KOI-5157.01 and 7430.01 and the multi-planet system consisting of KOIs 153.01 and 153.02. We note that all four of these KOIs have unusually poorly constrained transit epochs, the parameter that is generally best constrained by our modeling. Correspondingly, we recommend against adopting our modeled transit parameters for these KOIs and their host stars.

Our transit modeling-derived $\log_{10}\rho_*$ is more precise than the isochrone modeling-derived DR25 constraint for $50 \%$ of our targets. The median improvement to fractional uncertainty for these $50\%$ is a factor of 2.3, meaning that our fractional uncertainty is less than half that of the literature value for a typical target star. We achieve comparable precision to asteroseismology (i.e., fractional uncertainty in $\rho_* \leq 5 \%$) for $62 \%$ of our targets. For some others, e.g. KOI-1793.01, KOI-4351.01, and KOI-3913.01, the bimodality discussed in section~\ref{subsec:covariances} is apparent, and we derive a poor constraint on $\log_{10}\rho_*$.

We are able to extend sub-$5\%$-fractional-uncertainty stellar density measurements to \textit{Kepler} stars three magnitudes fainter than asteroseismology can, across a broad range in $T_{\mathrm{eff}}$, as shown in Figure~\ref{fig:target_selection}. Opaque circles in this figure represent stars for which we achieve fractional $\log_{10}\rho_*$ uncertainty of less than $5\%$, and transparent squares represent stars for which we do not.

%In Figure~\ref{fig:target_selection}, which shows our target list in $T_{\mathrm{eff}}$-magnitude space, we plot stars with published asteroseismic density measurements as black triangles, and stars for which we achieve asteroseismic-level precision on $\rho_*$ via transit modeling as colored triangles. 

\subsubsection{Comparisons with asteroseismology}\label{subsubsec:AS}

\begin{figure}[ht!]
\begin{center}
\includegraphics[width=0.45\textwidth]{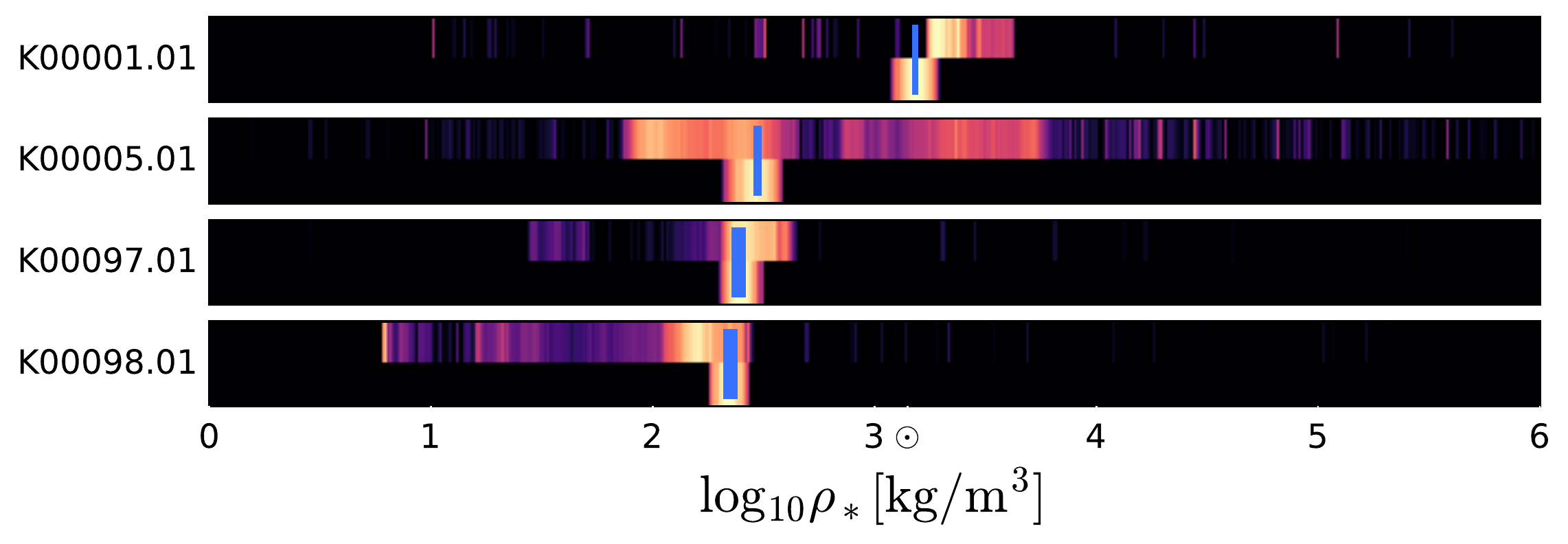}
\caption{A comparison of our $\rho_*$ posteriors (top rows) to the \cite{mathur17} $\rho_*$ posteriors (bottom rows) and asteroseismic constraints on $\rho_*$ by \cite{huber13} (blue boxes). We discuss the evident discrepancy between our results and asteroseismology for KOI-1.01 (TrES-2b) in Section~\ref{subsubsec:AS}.}
\label{fig:AScomp}
\end{center}
\end{figure}

Four of our occultation targets (KOIs 1.01, 5.01, 97.01, and 98.01) have previously been targeted for asteroseismic density measurement by \cite{huber13}. In Figure~\ref{fig:AScomp}, we compare our $\rho_*$ posteriors directly to the asteroseismically measured $\rho_*$ for each of these four targets.\footnote{The \citealt{mathur17} posteriors are derived by feeding these asteroseismic constraints through isochrone modeling, so they are in excellent agreement with the \citealt{huber13} results.}

Our results are in good agreement for KOIs 5.01, 97.01, and 98.01; KOIs 5.01 and 97.01 are in $1\sigma$ agreement with asteroseismology, and KOI-98.01 in $2\sigma$ agreement. We achieve comparable precision to asteroseismology except in the case of KOI 5.01, which undergoes grazing transits at low signal to noise and is subject to the parameter covariances discussed in section~\ref{subsec:covariances}.

For the final target with available asteroseismic data, KOI-1.01 (TrES-2b), we derive a higher stellar density than previously published constraints. Although our transit model for this planet is well-converged and a good match to the \textit{Kepler} data, we note that our best-fitting parameters conflict with earlier results from very reliable analyses---in particular, we derive an eccentricity $e = 0.2^{+0.16}_{-0.08}$, while radial velocity observations agree that this planet's orbit is consistent with being circular \citep{odonovan06,odonovan10,kipping11,clm12}. $e$ and $\rho_*$ are covariant, and the sense of the covariance is such that a too-high $e$ would indeed cause us to overestimate $\rho_*$. We must then explain why we derive such a high $e$, especially given that we impose a strong $e$ prior which should favor near-zero values of $e$.

We attribute our implausibly high $e$ to a failure of our transit model to accurately capture the limb-darkening behavior of KOI-1.01's host star. Upon closer inspection of KOI-1.01's posterior distributions, we observe that the posterior distributions of the limb-darkening coefficients are strange, particularly that of $\alpha_r$. While the vast majority of our target KOIs exhibit well-behaved $\alpha_r$ distributions like those of KOI-929.01 (see Figure~\ref{fig:K00929post}: the $\alpha_r$ distribution is broad and peaked in the middle of the allowable $\alpha_r$ range), KOI-1.01's is instead narrowly peaked at $\alpha_r = 0.01^{+0.07}_{-0.01}$. In other words, it abuts the lower boundary of our uniform prior on $\alpha_r$, which indicates that the limb-darkening behavior of our highest-likelihood model is unphysical.

\begin{figure}[!ht]
\begin{center}
\includegraphics[width=0.45\textwidth]{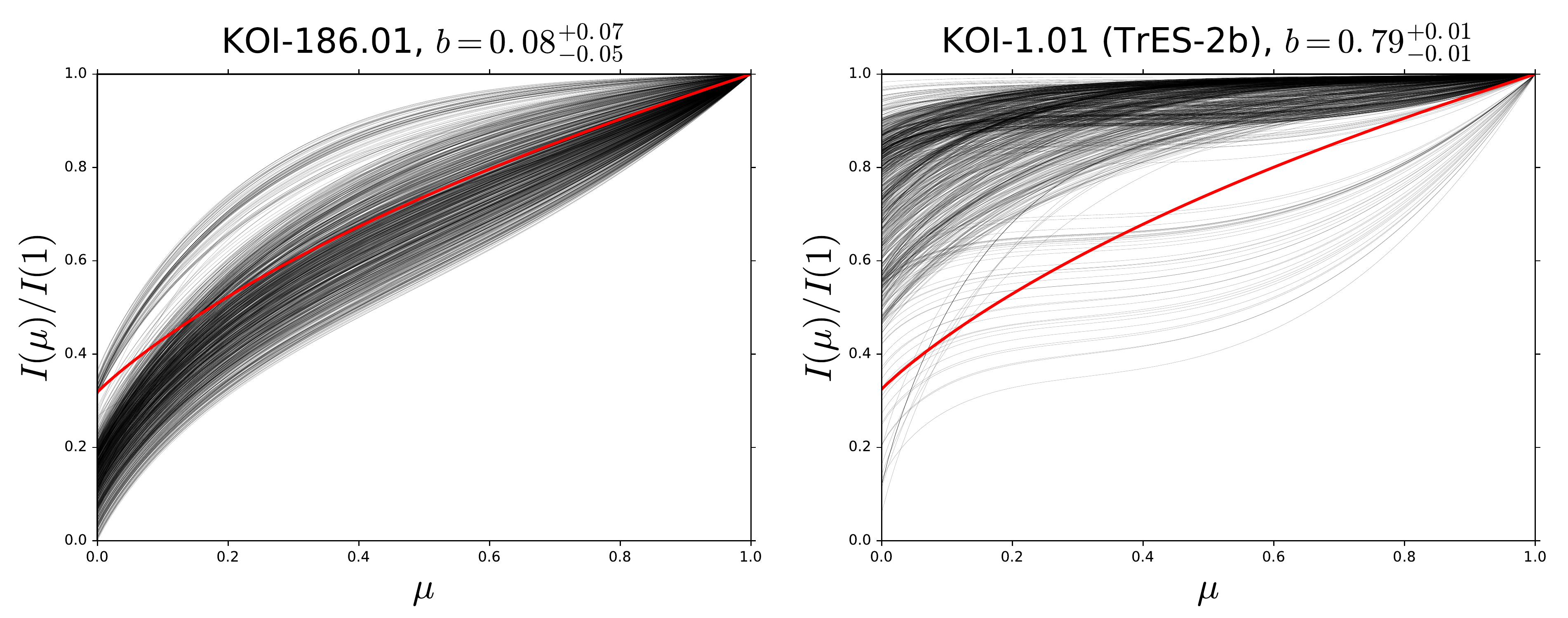}
\caption{The limb darkening profile of KOI-186.01, which transits close to the midplane of its host star, compared to that of KOI-1.01 (TrES-2b), which undergoes a grazing transit. In each panel, we plot profiles $I(\mu)/I(1) = 1 - c_2(1-\mu) - c_3(1-\mu^{3/2}) -c_4(1-\mu^2)$ computed from 1000 random draws from the posterior distributions of the LDCs (black lines) and theoretical predictions for the limb darkening profile from \cite{sing10} (red lines).}
\label{fig:grazingLD}
\end{center}
\end{figure}

Our transit model fails to capture the limb darkening behavior of KOI-1.01's host star because KOI-1.01 undergoes a grazing transit (we derive $b = 0.79\pm0.01$; the NEA reports $b = 0.818\pm0.001$). In other words, the planet transits across the limb of the star, so the transit data contains no information about the star's limb darkening behavior near the center of the sky-projected star ($\mu = 1$). Our three-parameter limb darkening law is actually somewhat of a liability in this situation: a very flexible model, subject to minimal constraining data, is free to adopt physically implausible (though still technically permitted within the bounds of the priors) combinations of the $\alpha$s in pursuit of the highest-likelihood solution, where a less flexible model, with fewer free parameters, would be fixed by fewer constraints.

To illustrate the undesirable effects of the flexibility in the limb darkening model in the case of grazing transits, in Figure~\ref{fig:grazingLD}, we compare our measured limb darkening profile of KOI-1.01 (TrES-2b) to that of KOI-186.01, which transits very close to the midplane of its host star. KOI-186.01's limb darkening profile is well-constrained from $\mu=0$ (the stellar limb) to $\mu=1$, while KOI-1.01's is poorly constrained, with a wide range of plausible $\alpha$ behavior.

Visual inspection of the posterior distributions for KOI-1.01 confirms that the posterior samples with near-zero $\alpha_r$ correspond to unrealistically high values of $e$ and $\rho_*$.

\begin{figure*}[!ht]
\begin{center}
\includegraphics[width=0.9\textwidth]{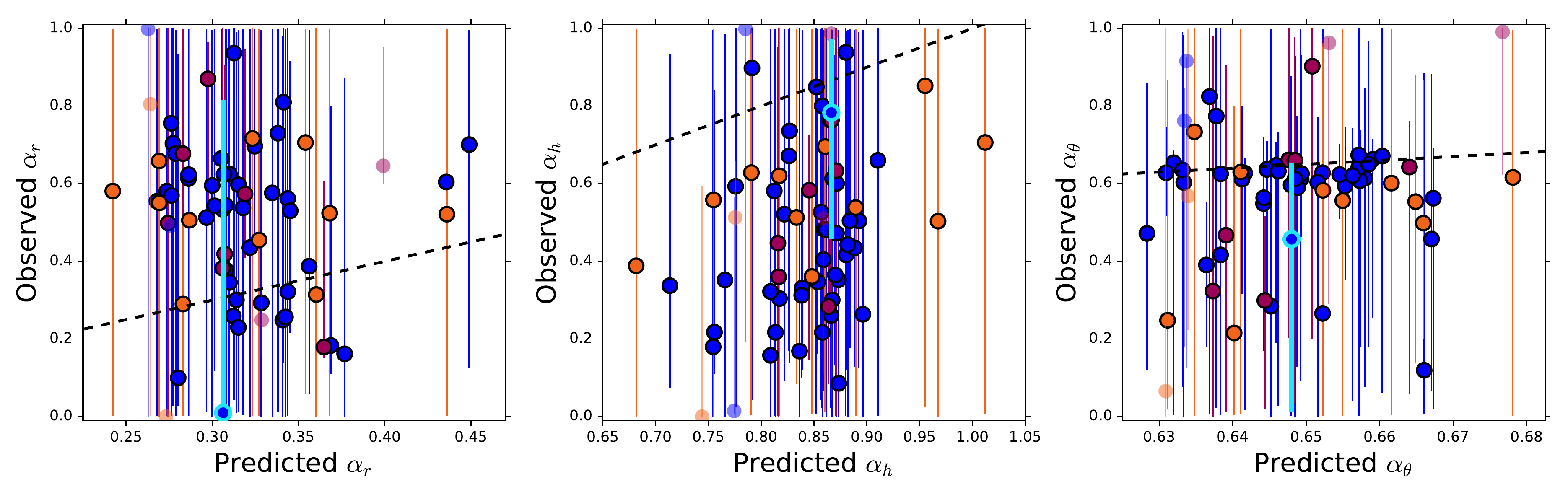}
\caption{A comparison of our $\alpha_r$, $\alpha_h$, and $\alpha_\theta$ to the theoretical predictions of \cite{sing10}, based on stellar atmosphere modeling. We plot $3\sigma$ uncertainty bands on our values to check for broad consistency. Blue points are occultation targets; orange points are tidal targets; purple points are multi-planet targets. The dotted lines indicate one-to-one correspondence, and KOI-1.01 (TrES-2b) is highlighted in bright blue. Transparent points are those with at least one anomalous $\alpha$ distribution, abutting either the upper or lower boundary of the prior.}
\label{fig:singLDCcomp}
\end{center}
\end{figure*}

We examine the remainder of our target list for other stars which exhibit similarly suspicious $\alpha$ posteriors, and we also compare our results to the theoretical predictions of \cite{sing10}, based on stellar atmosphere modeling, in figure~\ref{fig:singLDCcomp}. This figure shows a comparison of our observed $\alpha$-values to the \cite{sing10} predictions, which we calculate by linearly interpolating their Table 2 results and evaluating the interpolation at the NEA-provided stellar spectroscopic parameters for our target stars. To check for general consistency between our results and the \cite{sing10} predictions, we plot $3\sigma$ uncertainty bounds on our $\alpha$-values. We highlight the results for KOI-1.01 in bright blue; note in particular its anomalously low $\alpha_r$ value.

We identify a handful of suspect stars which exhibit similarly anomalous values of any of the three $\alpha$s, abutting either the lower ($\alpha = 0$) or upper ($\alpha = 1$) boundaries of our prior, and which also have derived values of $e$ which are inconsistent with their strong eccentricity prior. The KOIs meeting these criteria are, of the occultation targets, KOIs-1.01 (as discussed already) and 823.01; of the tidal targets, KOIs-1075.01 and 1658.01; and of the multi-planet targets, the KOI-153 and 1779 systems. All of these lone KOIs, and at least one KOI orbiting each of the suspect multi-planet targets, undergo grazing transits, so their behavior is overall consistent with the case of KOI-1.01, discussed above. (We note that occultation target KOI-1541.01 also has an anomalously high $\alpha_h$, but that its correspondingly high eccentricity is consistent with the priors in $e\cos{\omega}$ and $e\sin{\omega}$ adopted from \cite{clm12}, and also that it transits at very low impact parameter, so its transits contain information about its entire limb darkening profile.)

The case of KOI-1.01 demonstrates that strangely behaved LDC posterior distributions strongly indicate that other transit model parameters---especially $\rho_*$---may not be reliable. We therefore advise extreme caution in adopting our transit model parameter posteriors for these stars.

Aside from these isolated cases, which comprise $12\%$ of our target list, Figure~\ref{fig:singLDCcomp} establishes that our results are generally in good agreement with the predictions of \cite{sing10}. $79\%$ of our target stars are consistent with \cite{sing10} at the $3\sigma$ level in all three $\alpha$-dimensions, and $94\%$ in at least two of the three $\alpha$-dimensions. We note that, although our results statistically agree, there are systematic offsets between our $\alpha_r$ and $\alpha_h$ values and those of \cite{sing10}; in particular, we overpredict $\alpha_r$ and underpredict $\alpha_h$. These two parameters, however, as we discuss in Section~\ref{subsec:covariances}, are not independent---rather, they co-vary in exactly the sense observed in this figure, with high $\alpha_r$ corresponding to low $\alpha_h$. We find that the fractional uncertainty in the $\alpha$s is positively correlated with impact parameter $b$, consistent with the results for KOI-1.01.

%\begin{figure*}[!ht]
%\begin{center}
%\includegraphics[width=\textwidth]{K00001_100000bestfitBATMAN.pdf}
%\caption{Our transit fit for occultation target KOI-1.01, a confirmed planet. The black points are the 853 transits observed for this KOI, detrended and stacked, and the red lines are light curve models computed by \texttt{BATMAN} \citep{kreidberg15} from 500 random draws from our 10-dimensional transit parameter posterior distributions. This transit fit is well-converged and a good match to the data, but the transit-derived $\rho_*$ disagrees slightly with the asteroseismic $\rho_*$ measurement of \cite{huber13}.}
%\label{fig:K00001batman}
%\end{center}
%\end{figure*}

\subsection{Limb darkening coefficients}\label{subsec:LDC_results}

\begin{figure*}[!ht]
\begin{center}
\includegraphics[width=\textwidth]{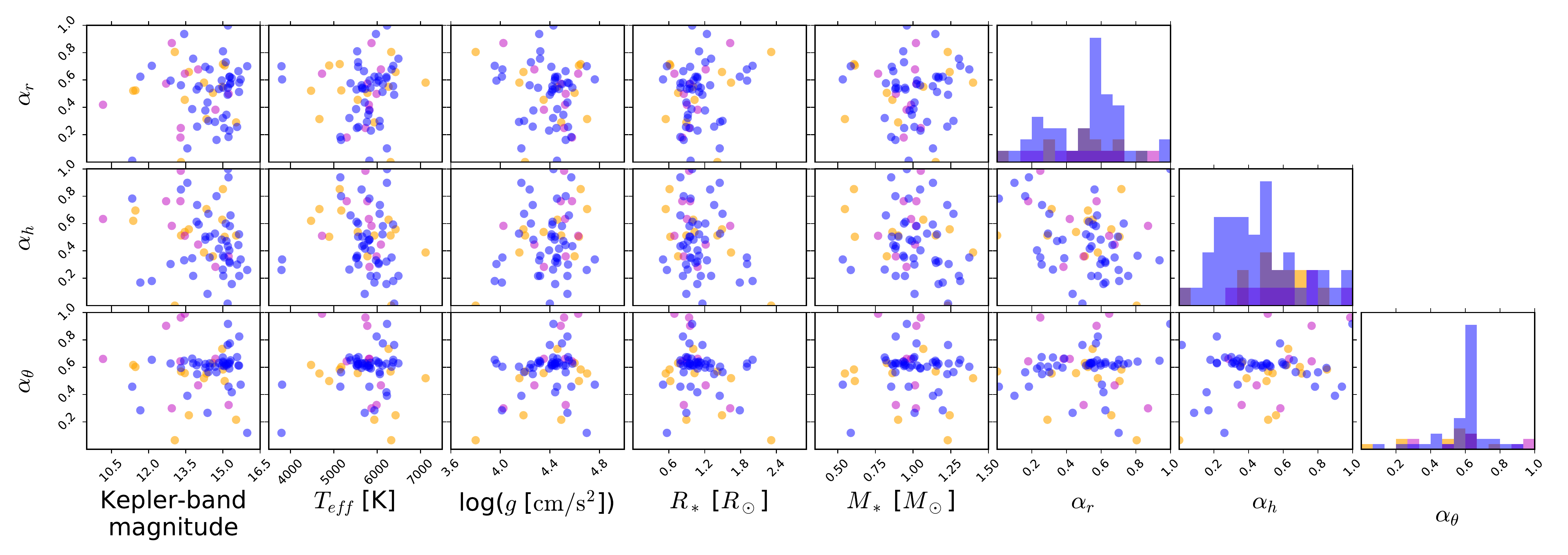}
\caption{Reparametrized limb darkening coefficients ($\alpha_r$, $\alpha_h$, and $\alpha_\theta$) vs. various properties of our target stars. Blue points are occultation targets; orange points are tidal targets; purple points are multi-planet targets. There are no significant correlations between the LDCs and stellar properties. We note that the $\alpha_\theta$ distribution is strongly peaked at 0.6 across a broad range of stellar properties, suggesting that this coefficient contains almost no information about the properties of the star.}
\label{fig:LDC}
\end{center}
\end{figure*}

We also investigate the relationship of our measured nonlinear limb-darkening coefficients, $\alpha_r$, $\alpha_h$, and $\alpha_\theta$, to various stellar properties from the \textit{Kepler} input catalog. In Figure~\ref{fig:LDC}, we plot various projections of this high-dimensional stellar parameter space to look for correlations. We find that the three $\alpha$s are totally uncorrelated with \textit{Kepler}-band magnitude, $T_{\mathrm{eff}}$, $\log{g}$, stellar radius, stellar mass, and each other. The only pattern of note in this parameter space is the sharp peak of $\alpha_\theta$ about a value of approximately 0.6; that so many disparate target stars share this value indicates that $\alpha_\theta$ is especially uninformative with respect to stellar properties. 

When we transform the three $\alpha$s into the more traditional nonlinear LDCs $c_2$, $c_3$, and $c_4$ (see e.g. \citealt{sing10,claret00}), the LDCs remain essentially uncorrelated with any stellar properties. The strong peak in the $\alpha_\theta$ distribution evident in Figure~\ref{fig:LDC} translates to a strong peak in $c_4$, indicating that $c_4$ is the least informative LDC with respect to stellar properties.

%In Figure~\ref{fig:transformedLDC}, we transform the three $\alpha$s into the more traditional nonlinear limb-darkening coefficients $c_2$, $c_3$, and $c_4$ (see e.g. \citealt{sing10,claret00}). These LDCs remain essentially uncorrelated with any stellar properties, but demonstrate strong correlations with each other; these strong correlations motivated the transformation to $\alpha$-space in the first place \citep{kipping16ldc}. The strong peak in the $\alpha_\theta$ distribution evident in Figure~\ref{fig:LDC} translates to a strong peak in $c_4$ here, indicating that $c_4$ is the least informative LDC with respect to stellar properties.

% perhaps not surprising given that it's the highest-order coefficient?

%\begin{figure*}[!ht]
%\begin{center}
%\includegraphics[width=\textwidth]{cropped_transformedLDC_scattercorner_horizontal.pdf}
%\caption{The traditional limb darkening coefficients $c_2$, $c_3$, and $c_4$ vs. various properties of our target stars. Blue points are occultation targets; orange points are tidal targets; purple points are multi-planet targets. There are no significant correlations between the LDCs and stellar properties; the correlations between the LDCs themselves, evident in the rightmost panels, motivated the reparametrization to $\alpha$-space originally \citep{kipping16ldc}. The strong peak of the $\alpha_\theta$ distribution in Figure~\ref{fig:LDC} translates to a strong peak in $c_4$ here, suggesting that $c_4$ is the least informative LDC with regard to stellar properties.}
%\label{fig:transformedLDC}
%\end{center}
%\end{figure*}

\subsubsection{Which stars are the best transit-modeling targets?} \label{sec:discussion}

Having demonstrated the capability of transit modeling to yield high-precision measurements of stellar density, we now ask: Are some stars better suited to measurement by this method, and if so, can we identify those stars ahead of time? In other words, are there any properties of a star or its associated KOIs that predict a successful, precise transit-based $\rho_*$ measurement, or disqualify a star from such a measurement?

\begin{figure*}[!ht]
\begin{center}
\includegraphics[width=\textwidth]{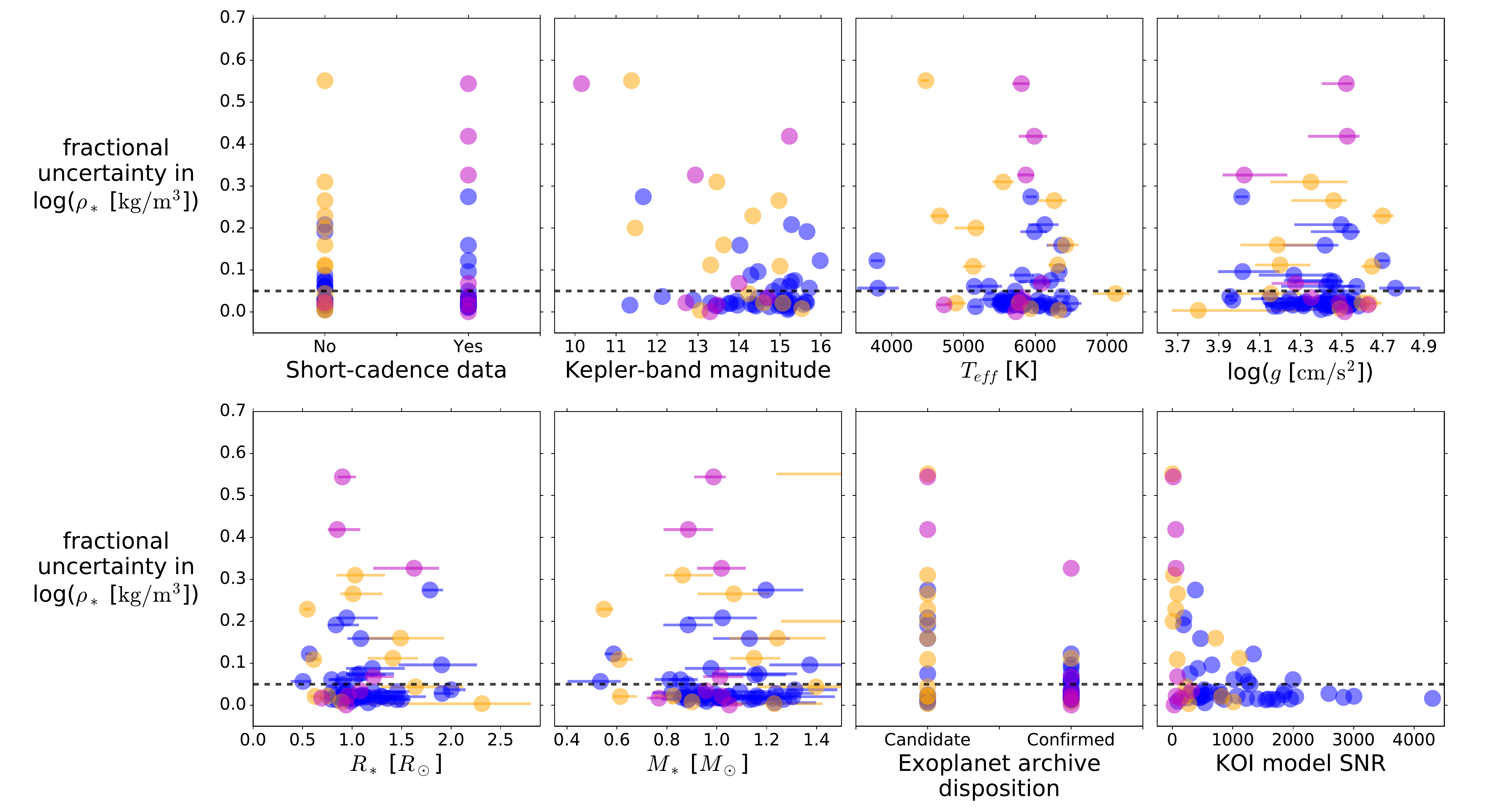}
\caption{The fractional uncertainty in $\rho_*$ vs. stellar and KOI data properties for each of our 66 targets. Blue points are occultation targets; orange points are tidal targets; purple points are multi-planet targets; the dotted lines in each panel mark $5 \%$ fractional uncertainty in $\rho_*$.}
\label{fig:rhostarunc}
\end{center}
\end{figure*}

In Figure~\ref{fig:rhostarunc}, we plot the fractional uncertainty of each of our transit-based $\rho_*$ measurements against various stellar and KOI data properties. Stellar properties include the \textit{Kepler}-band magnitude, $T_{\mathrm{eff}}$, $\log{g}$, stellar radius, and stellar mass; KOI properties include whether short-cadence observations were available, whether the KOI is ``confirmed" per its NEA disposition, and the NEA threshold-crossing event signal-to-noise ratio (SNR). 

We find that the achieved precision on $\rho_*$ does not depend on any stellar properties, meaning that faint and bright, hot and cool, large and small stars are equally appropriate targets, \textit{a priori}, for transit-based stellar density measurements. This lack of any dependence of the success of our method on the properties of our target stars is evident in the distribution of opaque circles (target stars measured to high $\rho_*$ precision) across Figure~\ref{fig:target_selection}.

The fractional uncertainty in our $\rho_*$ measurements does, however, correlate strongly with the NEA-reported SNR, which is sensible given that the precision of our derived transit parameters, including $\rho_*$, depends on our ability to successfully model transits. We furthermore find that planets observed in short cadence are more likely to have precise $\rho_*$ measurements, but that short-cadence data is not necessary to achieve this level of precision. Planets which are confirmed per the NEA are also more likely to have high-precision $\rho_*$ measurements, which is partly due to our ability to discard posterior samples with unphysical $R_p/R_*$ (and corresponding $\rho_*$) for these KOIs. Another contributing factor is that planets which are easy to confirm by other exoplanet detection methods (large, close to their host stars) are also likely to have high transit SNR.

\section{Conclusions}\label{sec:conc}
In this work, we demonstrate the promise of exoplanetary transits to characterize planet-hosting stars with high precision. We select 66 target planet-star systems with strong prior constraints on planetary eccentricity, either directly measured from secondary eclipses, or strongly implied by a short tidal circularization timescale or compact multiplicity. We fit transit models to these targets and derive posterior distributions of ten transit parameters: transit epoch, period, impact parameter, stellar density, ratio-of-radii, reparametrized eccentricity and argument of periastron, and three reparametrized coefficients of a modified nonlinear limb darkening law. We make downsampled posterior distributions for the transit parameters of the 75 KOIs orbiting our 66 target stars available at  \url{https://doi.org/10.5281/zenodo.1028515}.

For $95\%$ of our targets, our measured stellar densities are in agreement with previously published constraints at the $3\sigma$ level ($55\%$ at the $1\sigma$ level). Furthermore, for $50\%$ of our targets, we improve upon the best available published constraint on stellar density; the median improvement is slightly greater than a factor of two, meaning that we achieve a fractional uncertainty less than half of the literature value. For $62\%$ of our targets, we achieve comparable precision to asteroseismology (typical fractional uncertainty $\leq 5\%$), generally considered the gold-standard method of stellar density measurement. We demonstrate that the success of our method for any given target planet-star system does not depend on any of the star's properties, but instead depends only on the signal-to-noise ratio of the planetary transits. 

Correspondingly, we successfully use this method to extend asteroseismic-level-precision stellar density measurements to stars three magnitudes fainter than the \textit{Kepler} asteroseismic limit, across a broad range of effective temperatures. We note that, although TESS will observe brighter stars than \textit{Kepler}, its asteroseismic limit will be several magnitudes brighter ($\sim 8^{\textrm{th}}$ magnitude) due to its small aperture \citep{campante16, ricker14}, and therefore that this transit-based method will be invaluable in characterizing TESS stars which are inaccessible to asteroseismology.

We emphasize that this method requires no data beyond a transiting exoplanet light curve, and it therefore promises to aid greatly in exoplanet host star characterization in the era of TESS and LSST, when we expect to discover many more transiting planets than we can hope to quickly follow up spectroscopically. 

Finally, in advance of TESS, we note the potential of this precise transit fitting technique to characterize not just stars, but also singly-transiting planets, as shown by \citealt{yee08}. We demonstrate in this work that transiting planets with strong prior eccentricity constraints may be used to strongly constrain their host stars, yielding very precise ($\leq 5\%$ uncertainty) measurements of their hosts' properties, including $\rho_*$. Once a host star is ``anchored" by a transiting planet (a ``\textit{stellar anchor}" planet) in this way, the properties of any \textit{other} transiting planets in the system can be derived to higher precision using the transit-measured stellar parameters than would be possible without this information. 

\begin{figure}[!ht]
\begin{center}
\includegraphics[width=0.48\textwidth]{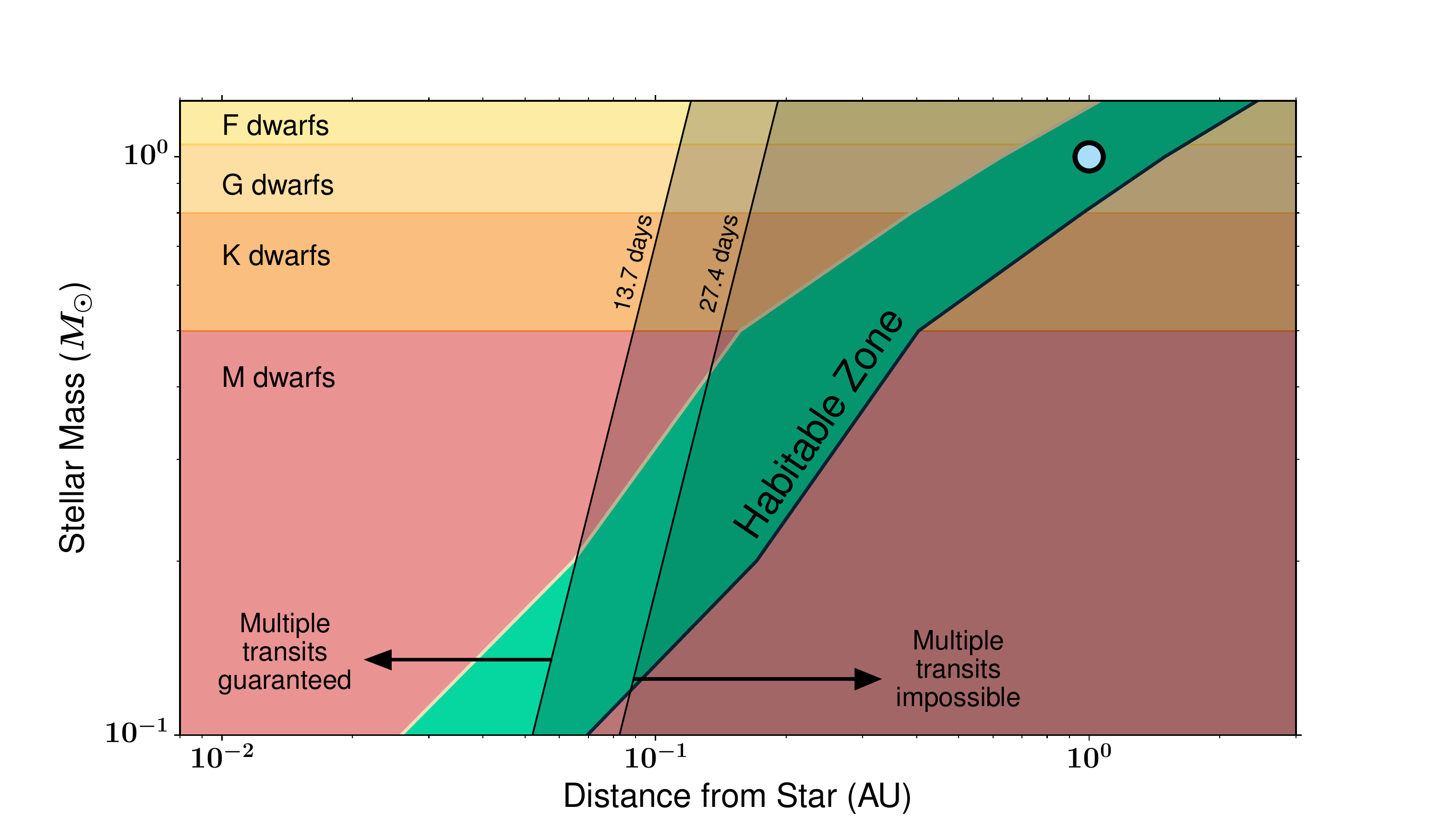}
\caption{The approximate location of the habitable zone about late-type stars \citep{kopparapu13}. Planets in the dark gray shaded region will transit at most once during TESS' 27.4-day single-visit observational baseline. Earth is plotted as a blue dot.}
\label{fig:HZ}
\end{center}
\end{figure}

TESS' observational baseline will be only 27.4 days over $\sim 30,000\, \mathrm{deg}^2$ on the sky; in these regions, TESS will be unable to directly measure the period of any planet with a period greater than 27.4 days, because it will observe at most one transit of such a planet \citep{ricker14}. As shown in Figure~\ref{fig:HZ}, this short baseline precludes direct measurement of the periods of planets in a large region of ``habitable zone" parameter space.

However, if any of these long-period, singly-transiting planets orbits the same star as a short-period stellar anchor planet, we will be able to use the anchor's transits to precisely measure $\rho_*$ with the method developed in this work, then place better constraints on the period of the single-transiter using the stellar density constraint. 

In the simplest case of an outer single transiter on a circular orbit, we expect the fractional uncertainty of the period $P$ to equal
\begin{equation}
    \frac{\sigma_P}{P} = \frac{1}{2} \sqrt{\left(\frac{\sigma_{\rho_*}}{\rho_*}\right)^2 - \left(3\frac{ \sigma_{(a/R_*)}}{a/R_*}\right)^2}
\end{equation}

by the propagation of uncertainty through equation~\ref{eq:rhostar}. Here, $P$ and $a/R_*$ are parameters of the outer single transiter, and $\rho_*$ is the density of the star that the outer single transiter and the inner stellar anchor both orbit.

We assume that the fractional uncertainty in the stellar density, $\frac{\sigma_{\rho_*}}{\rho_*}$, will dominate over the fractional uncertainty in the normalized semimajor axis, $\frac{\sigma_{(a/R_*)}}{a/R_*}$, because $a/R_*$ can be measured directly from the shape of the single observed transit of the outer planet, while $\rho_*$ requires detailed modeling of the inner stellar anchor's transits subject to a prior constraint on eccentricity, as described in this work. Under this assumption, a $5\%$ fractional uncertainty in $\rho_*$ corresponds to a $2.5\%$ fractional uncertainty in the period $P$ of the single transiter. 

While considerably less precise than a direct period measurement, this degree of fractional uncertainly could establish whether a given planet orbits in the habitable zone of its host star or not, and hence whether it merits follow-up observations. Obtaining a similar constraint on the period using radial velocity measurements of the host star would be time-consuming, generally requiring an observational baseline comparable to the orbital period of the planet (see e.g. \citealt{ford05}). Our method therefore promises to aid greatly in the characterization of long-period TESS planets.\\

\acknowledgments

The authors thank members of the Cool Worlds Lab for useful discussions. This research has made use of the NASA Exoplanet Archive, which is operated by the California Institute of Technology, under contract with the National Aeronautics and Space Administration under the Exoplanet Exploration Program.
\\
\software{\texttt{BATMAN} \citep{kreidberg15}, \texttt{emcee} \citep{dfm13}, \texttt{corner.py} \citep{dfm16}, \texttt{scipy} \citep{scipy}, and \texttt{astropy} \citep{astropy}.}

\bibliography{transitsbib}

\begin{thebibliography}{}
\expandafter\ifx\csname natexlab\endcsname\relax\def\natexlab#1{#1}\fi

\bibitem[{{Aigrain} {et~al.}(2016){Aigrain}, {Parviainen}, \&
  {Pope}}]{aigrain16}
{Aigrain}, S., {Parviainen}, H., \& {Pope}, B.~J.~S. 2016, \mnras, 459, 2408

\bibitem[{{Akeson}(2015)}]{akeson15}
{Akeson}, R.~L. 2015, in AAS/Division for Planetary Sciences Meeting Abstracts,
  Vol.~47, AAS/Division for Planetary Sciences Meeting Abstracts, 417.01

\bibitem[{Anderson \& Darling(1952)}]{ad52}
Anderson, T.~W., \& Darling, D.~A. 1952, Ann. Math. Statist., 23, 193

\bibitem[{{Astropy Collaboration} {et~al.}(2013){Astropy Collaboration},
  {Robitaille}, {Tollerud}, {Greenfield}, {Droettboom}, {Bray}, {Aldcroft},
  {Davis}, {Ginsburg}, {Price-Whelan}, {Kerzendorf}, {Conley}, {Crighton},
  {Barbary}, {Muna}, {Ferguson}, {Grollier}, {Parikh}, {Nair}, {Unther},
  {Deil}, {Woillez}, {Conseil}, {Kramer}, {Turner}, {Singer}, {Fox}, {Weaver},
  {Zabalza}, {Edwards}, {Azalee Bostroem}, {Burke}, {Casey}, {Crawford},
  {Dencheva}, {Ely}, {Jenness}, {Labrie}, {Lim}, {Pierfederici}, {Pontzen},
  {Ptak}, {Refsdal}, {Servillat}, \& {Streicher}}]{astropy}
{Astropy Collaboration}, {Robitaille}, T.~P., {Tollerud}, E.~J., {et~al.} 2013,
  \aap, 558, A33

\bibitem[{{Batalha} {et~al.}(2013){Batalha}, {Rowe}, {Bryson}, {Barclay},
  {Burke}, {Caldwell}, {Christiansen}, {Mullally}, {Thompson}, {Brown},
  {Dupree}, {Fabrycky}, {Ford}, {Fortney}, {Gilliland}, {Isaacson}, {Latham},
  {Marcy}, {Quinn}, {Ragozzine}, {Shporer}, {Borucki}, {Ciardi}, {Gautier},
  {Haas}, {Jenkins}, {Koch}, {Lissauer}, {Rapin}, {Basri}, {Boss}, {Buchhave},
  {Carter}, {Charbonneau}, {Christensen-Dalsgaard}, {Clarke}, {Cochran},
  {Demory}, {Desert}, {Devore}, {Doyle}, {Esquerdo}, {Everett}, {Fressin},
  {Geary}, {Girouard}, {Gould}, {Hall}, {Holman}, {Howard}, {Howell},
  {Ibrahim}, {Kinemuchi}, {Kjeldsen}, {Klaus}, {Li}, {Lucas}, {Meibom},
  {Morris}, {Pr{\v s}a}, {Quintana}, {Sanderfer}, {Sasselov}, {Seader},
  {Smith}, {Steffen}, {Still}, {Stumpe}, {Tarter}, {Tenenbaum}, {Torres},
  {Twicken}, {Uddin}, {Van Cleve}, {Walkowicz}, \& {Welsh}}]{batalha13}
{Batalha}, N.~M., {Rowe}, J.~F., {Bryson}, S.~T., {et~al.} 2013, \apjs, 204, 24

\bibitem[{{Campante} {et~al.}(2016){Campante}, {Schofield}, {Kuszlewicz},
  {Bouma}, {Chaplin}, {Huber}, {Christensen-Dalsgaard}, {Kjeldsen}, {Bossini},
  {North}, {Appourchaux}, {Latham}, {Pepper}, {Ricker}, {Stassun},
  {Vanderspek}, \& {Winn}}]{campante16}
{Campante}, T.~L., {Schofield}, M., {Kuszlewicz}, J.~S., {et~al.} 2016, \apj,
  830, 138

\bibitem[{{Carter} {et~al.}(2008){Carter}, {Yee}, {Eastman}, {Gaudi}, \&
  {Winn}}]{carter08}
{Carter}, J.~A., {Yee}, J.~C., {Eastman}, J., {Gaudi}, B.~S., \& {Winn}, J.~N.
  2008, \apj, 689, 499

\bibitem[{{Claret}(2000)}]{claret00}
{Claret}, A. 2000, \aap, 363, 1081

\bibitem[{{Coughlin} \& {L{\'o}pez-Morales}(2012)}]{clm12}
{Coughlin}, J.~L., \& {L{\'o}pez-Morales}, M. 2012, \aj, 143, 39

\bibitem[{{D{\'e}sert} {et~al.}(2011){D{\'e}sert}, {Charbonneau}, {Demory},
  {Ballard}, {Carter}, {Fortney}, {Cochran}, {Endl}, {Quinn}, {Isaacson},
  {Fressin}, {Buchhave}, {Latham}, {Knutson}, {Bryson}, {Torres}, {Rowe},
  {Batalha}, {Borucki}, {Brown}, {Caldwell}, {Christiansen}, {Deming},
  {Fabrycky}, {Ford}, {Gilliland}, {Gillon}, {Haas}, {Jenkins}, {Kinemuchi},
  {Koch}, {Lissauer}, {Lucas}, {Mullally}, {MacQueen}, {Marcy}, {Sasselov},
  {Seager}, {Still}, {Tenenbaum}, {Uddin}, \& {Winn}}]{desert11}
{D{\'e}sert}, J.-M., {Charbonneau}, D., {Demory}, B.-O., {et~al.} 2011, \apjs,
  197, 14

\bibitem[{{Espinoza} \& {Jord{\'a}n}(2015)}]{espinoza15}
{Espinoza}, N., \& {Jord{\'a}n}, A. 2015, \mnras, 450, 1879

\bibitem[{{Fabrycky} {et~al.}(2012){Fabrycky}, {Ford}, {Steffen}, {Rowe},
  {Carter}, {Moorhead}, {Batalha}, {Borucki}, {Bryson}, {Buchhave},
  {Christiansen}, {Ciardi}, {Cochran}, {Endl}, {Fanelli}, {Fischer}, {Fressin},
  {Geary}, {Haas}, {Hall}, {Holman}, {Jenkins}, {Koch}, {Latham}, {Li},
  {Lissauer}, {Lucas}, {Marcy}, {Mazeh}, {McCauliff}, {Quinn}, {Ragozzine},
  {Sasselov}, \& {Shporer}}]{fabrycky12}
{Fabrycky}, D.~C., {Ford}, E.~B., {Steffen}, J.~H., {et~al.} 2012, \apj, 750,
  114

\bibitem[{{Ford}(2005)}]{ford05}
{Ford}, E.~B. 2005, \aj, 129, 1706

\bibitem[{Foreman-Mackey(2016)}]{dfm16}
Foreman-Mackey, D. 2016, The Journal of Open Source Software, 24,
  doi:10.21105/joss.00024

\bibitem[{{Foreman-Mackey} {et~al.}(2013){Foreman-Mackey}, {Hogg}, {Lang}, \&
  {Goodman}}]{dfm13}
{Foreman-Mackey}, D., {Hogg}, D.~W., {Lang}, D., \& {Goodman}, J. 2013, \pasp,
  125, 306

\bibitem[{{Haswell}(2010)}]{haswell10}
{Haswell}, C.~A. 2010, {Transiting Exoplanets}

\bibitem[{{Holczer} {et~al.}(2016){Holczer}, {Mazeh}, {Nachmani},
  {Jontof-Hutter}, {Ford}, {Fabrycky}, {Ragozzine}, {Kane}, \&
  {Steffen}}]{holczer16}
{Holczer}, T., {Mazeh}, T., {Nachmani}, G., {et~al.} 2016, \apjs, 225, 9

\bibitem[{{Huber} {et~al.}(2013){Huber}, {Chaplin}, {Christensen-Dalsgaard},
  {Gilliland}, {Kjeldsen}, {Buchhave}, {Fischer}, {Lissauer}, {Rowe},
  {Sanchis-Ojeda}, {Basu}, {Handberg}, {Hekker}, {Howard}, {Isaacson},
  {Karoff}, {Latham}, {Lund}, {Lundkvist}, {Marcy}, {Miglio}, {Silva Aguirre},
  {Stello}, {Arentoft}, {Barclay}, {Bedding}, {Burke}, {Christiansen},
  {Elsworth}, {Haas}, {Kawaler}, {Metcalfe}, {Mullally}, \&
  {Thompson}}]{huber13}
{Huber}, D., {Chaplin}, W.~J., {Christensen-Dalsgaard}, J., {et~al.} 2013,
  \apj, 767, 127

\bibitem[{{Ivezic} {et~al.}(2008){Ivezic}, {Tyson}, {Abel}, {Acosta},
  {Allsman}, {AlSayyad}, {Anderson}, {Andrew}, {Angel}, {Angeli}, {Ansari},
  {Antilogus}, {Arndt}, {Astier}, {Aubourg}, {Axelrod}, {Bard}, {Barr},
  {Barrau}, {Bartlett}, {Bauman}, {Beaumont}, {Becker}, {Becla}, {Beldica},
  {Bellavia}, {Blanc}, {Blandford}, {Bloom}, {Bogart}, {Borne}, {Bosch},
  {Boutigny}, {Brandt}, {Brown}, {Bullock}, {Burchat}, {Burke}, {Cagnoli},
  {Calabrese}, {Chandrasekharan}, {Chesley}, {Cheu}, {Chiang}, {Claver},
  {Connolly}, {Cook}, {Cooray}, {Covey}, {Cribbs}, {Cui}, {Cutri}, {Daubard},
  {Daues}, {Delgado}, {Digel}, {Doherty}, {Dubois}, {Dubois-Felsmann},
  {Durech}, {Eracleous}, {Ferguson}, {Frank}, {Freemon}, {Gangler}, {Gawiser},
  {Geary}, {Gee}, {Geha}, {Gibson}, {Gilmore}, {Glanzman}, {Goodenow},
  {Gressler}, {Gris}, {Guyonnet}, {Hascall}, {Haupt}, {Hernandez}, {Hogan},
  {Huang}, {Huffer}, {Innes}, {Jacoby}, {Jain}, {Jee}, {Jernigan},
  {Jevremovic}, {Johns}, {Jones}, {Juramy-Gilles}, {Juric}, {Kahn}, {Kalirai},
  {Kallivayalil}, {Kalmbach}, {Kantor}, {Kasliwal}, {Kessler}, {Kirkby},
  {Knox}, {Kotov}, {Krabbendam}, {Krughoff}, {Kubanek}, {Kuczewski},
  {Kulkarni}, {Lambert}, {Le Guillou}, {Levine}, {Liang}, {Lim}, {Lintott},
  {Lupton}, {Mahabal}, {Marshall}, {Marshall}, {May}, {McKercher}, {Migliore},
  {Miller}, {Mills}, {Monet}, {Moniez}, {Neill}, {Nief}, {Nomerotski},
  {Nordby}, {O'Connor}, {Oliver}, {Olivier}, {Olsen}, {Ortiz}, {Owen}, {Pain},
  {Peterson}, {Petry}, {Pierfederici}, {Pietrowicz}, {Pike}, {Pinto}, {Plante},
  {Plate}, {Price}, {Prouza}, {Radeka}, {Rajagopal}, {Rasmussen}, {Regnault},
  {Ridgway}, {Ritz}, {Rosing}, {Roucelle}, {Rumore}, {Russo}, {Saha},
  {Sassolas}, {Schalk}, {Schindler}, {Schneider}, {Schumacher}, {Sebag},
  {Sembroski}, {Seppala}, {Shipsey}, {Silvestri}, {Smith}, {Smith}, {Strauss},
  {Stubbs}, {Sweeney}, {Szalay}, {Takacs}, {Thaler}, {Van Berg}, {Vanden Berk},
  {Vetter}, {Virieux}, {Xin}, {Walkowicz}, {Walter}, {Wang}, {Warner},
  {Willman}, {Wittman}, {Wolff}, {Wood-Vasey}, {Yoachim}, {Zhan}, \& {for the
  LSST Collaboration}}]{lsst}
{Ivezic}, Z., {Tyson}, J.~A., {Abel}, B., {et~al.} 2008, ArXiv e-prints,
  arXiv:0805.2366

\bibitem[{Jones {et~al.}(2001--)Jones, Oliphant, Peterson, {et~al.}}]{scipy}
Jones, E., Oliphant, T., Peterson, P., {et~al.} 2001--, {SciPy}: Open source
  scientific tools for {Python}, ,

\bibitem[{{Kipping} \& {Bakos}(2011)}]{kipping11}
{Kipping}, D., \& {Bakos}, G. 2011, \apj, 733, 36

\bibitem[{{Kipping}(2008)}]{kipping08}
{Kipping}, D.~M. 2008, \mnras, 389, 1383

\bibitem[{{Kipping}(2010)}]{kipping10}
---. 2010, \mnras, 407, 301

\bibitem[{{Kipping}(2013)}]{kipping13}
---. 2013, \mnras, 434, L51

\bibitem[{{Kipping}(2016)}]{kipping16ldc}
---. 2016, \mnras, 455, 1680

\bibitem[{{Kipping} \& {Sandford}(2016)}]{kippingsandford16}
{Kipping}, D.~M., \& {Sandford}, E. 2016, \mnras, 463, 1323

\bibitem[{{Knutson} {et~al.}(2007{\natexlab{a}}){Knutson}, {Charbonneau},
  {Noyes}, {Brown}, \& {Gilliland}}]{knutson07}
{Knutson}, H.~A., {Charbonneau}, D., {Noyes}, R.~W., {Brown}, T.~M., \&
  {Gilliland}, R.~L. 2007{\natexlab{a}}, \apj, 655, 564

\bibitem[{{Knutson} {et~al.}(2007{\natexlab{b}}){Knutson}, {Charbonneau},
  {Allen}, {Fortney}, {Agol}, {Cowan}, {Showman}, {Cooper}, \&
  {Megeath}}]{knutson07nat}
{Knutson}, H.~A., {Charbonneau}, D., {Allen}, L.~E., {et~al.}
  2007{\natexlab{b}}, \nat, 447, 183

\bibitem[{Kolmogorov(1933)}]{k33}
Kolmogorov, A.~N. 1933, Giornale dell'Istituto Italiano degli Attuari, 4, 83

\bibitem[{{Kopparapu} {et~al.}(2013){Kopparapu}, {Ramirez}, {Kasting}, {Eymet},
  {Robinson}, {Mahadevan}, {Terrien}, {Domagal-Goldman}, {Meadows}, \&
  {Deshpande}}]{kopparapu13}
{Kopparapu}, R.~K., {Ramirez}, R., {Kasting}, J.~F., {et~al.} 2013, \apj, 765,
  131

\bibitem[{{Kreidberg}(2015)}]{kreidberg15}
{Kreidberg}, L. 2015, \pasp, 127, 1161

\bibitem[{{Kundurthy} {et~al.}(2011){Kundurthy}, {Agol}, {Becker}, {Barnes},
  {Williams}, \& {Mukadam}}]{kundurthy11}
{Kundurthy}, P., {Agol}, E., {Becker}, A.~C., {et~al.} 2011, \apj, 731, 123

\bibitem[{{Lissauer} {et~al.}(2011){Lissauer}, {Fabrycky}, {Ford}, {Borucki},
  {Fressin}, {Marcy}, {Orosz}, {Rowe}, {Torres}, {Welsh}, {Batalha}, {Bryson},
  {Buchhave}, {Caldwell}, {Carter}, {Charbonneau}, {Christiansen}, {Cochran},
  {Desert}, {Dunham}, {Fanelli}, {Fortney}, {Gautier}, {Geary}, {Gilliland},
  {Haas}, {Hall}, {Holman}, {Koch}, {Latham}, {Lopez}, {McCauliff}, {Miller},
  {Morehead}, {Quintana}, {Ragozzine}, {Sasselov}, {Short}, \&
  {Steffen}}]{lissauer11}
{Lissauer}, J.~J., {Fabrycky}, D.~C., {Ford}, E.~B., {et~al.} 2011, \nat, 470,
  53

\bibitem[{Love(1934)}]{love34}
Love, A. 1934, A Treatise on the Mathematical Theory of Elasticity (University
  Press)

\bibitem[{{Luger} {et~al.}(2016){Luger}, {Agol}, {Kruse}, {Barnes}, {Becker},
  {Foreman-Mackey}, \& {Deming}}]{luger16}
{Luger}, R., {Agol}, E., {Kruse}, E., {et~al.} 2016, \aj, 152, 100

\bibitem[{{Mathur} {et~al.}(2017){Mathur}, {Huber}, {Batalha}, {Ciardi},
  {Bastien}, {Bieryla}, {Buchhave}, {Cochran}, {Endl}, {Esquerdo}, {Furlan},
  {Howard}, {Howell}, {Isaacson}, {Latham}, {MacQueen}, \& {Silva}}]{mathur17}
{Mathur}, S., {Huber}, D., {Batalha}, N.~M., {et~al.} 2017, \apjs, 229, 30

\bibitem[{{Morton} {et~al.}(2016){Morton}, {Bryson}, {Coughlin}, {Rowe},
  {Ravichandran}, {Petigura}, {Haas}, \& {Batalha}}]{morton16}
{Morton}, T.~D., {Bryson}, S.~T., {Coughlin}, J.~L., {et~al.} 2016, \apj, 822,
  86

\bibitem[{{O'Donovan} {et~al.}(2010){O'Donovan}, {Charbonneau}, {Harrington},
  {Madhusudhan}, {Seager}, {Deming}, \& {Knutson}}]{odonovan10}
{O'Donovan}, F.~T., {Charbonneau}, D., {Harrington}, J., {et~al.} 2010, \apj,
  710, 1551

\bibitem[{{O'Donovan} {et~al.}(2006){O'Donovan}, {Charbonneau}, {Mandushev},
  {Dunham}, {Latham}, {Torres}, {Sozzetti}, {Brown}, {Trauger}, {Belmonte},
  {Rabus}, {Almenara}, {Alonso}, {Deeg}, {Esquerdo}, {Falco}, {Hillenbrand},
  {Roussanova}, {Stefanik}, \& {Winn}}]{odonovan06}
{O'Donovan}, F.~T., {Charbonneau}, D., {Mandushev}, G., {et~al.} 2006, \apjl,
  651, L61

\bibitem[{{Ogilvie} \& {Lin}(2004)}]{ogilvie04}
{Ogilvie}, G.~I., \& {Lin}, D.~N.~C. 2004, \apj, 610, 477

\bibitem[{{O'Leary} \& {Burkart}(2014)}]{oleary14}
{O'Leary}, R.~M., \& {Burkart}, J. 2014, \mnras, 440, 3036

\bibitem[{{Orosz} {et~al.}(2012){Orosz}, {Welsh}, {Carter}, {Fabrycky},
  {Cochran}, {Endl}, {Ford}, {Haghighipour}, {MacQueen}, {Mazeh},
  {Sanchis-Ojeda}, {Short}, {Torres}, {Agol}, {Buchhave}, {Doyle}, {Isaacson},
  {Lissauer}, {Marcy}, {Shporer}, {Windmiller}, {Barclay}, {Boss}, {Clarke},
  {Fortney}, {Geary}, {Holman}, {Huber}, {Jenkins}, {Kinemuchi}, {Kruse},
  {Ragozzine}, {Sasselov}, {Still}, {Tenenbaum}, {Uddin}, {Winn}, {Koch}, \&
  {Borucki}}]{orosz12}
{Orosz}, J.~A., {Welsh}, W.~F., {Carter}, J.~A., {et~al.} 2012, Science, 337,
  1511

\bibitem[{Poulsen(2009)}]{poulsen09}
Poulsen, S.~K. 2009, Master's thesis, Niels Bohr Institute, University of
  Copenhagen

\bibitem[{{Rappaport} {et~al.}(2016){Rappaport}, {Lehmann}, {Kalomeni},
  {Borkovits}, {Latham}, {Bieryla}, {Ngo}, {Mawet}, {Howell}, {Horch},
  {Jacobs}, {LaCourse}, {S{\'o}dor}, {Vanderburg}, \&
  {Pavlovski}}]{rappaport16}
{Rappaport}, S., {Lehmann}, H., {Kalomeni}, B., {et~al.} 2016, \mnras, 462,
  1812

\bibitem[{{Ricker} {et~al.}(2014){Ricker}, {Winn}, {Vanderspek}, {Latham},
  {Bakos}, {Bean}, {Berta-Thompson}, {Brown}, {Buchhave}, {Butler}, {Butler},
  {Chaplin}, {Charbonneau}, {Christensen-Dalsgaard}, {Clampin}, {Deming},
  {Doty}, {De Lee}, {Dressing}, {Dunham}, {Endl}, {Fressin}, {Ge}, {Henning},
  {Holman}, {Howard}, {Ida}, {Jenkins}, {Jernigan}, {Johnson}, {Kaltenegger},
  {Kawai}, {Kjeldsen}, {Laughlin}, {Levine}, {Lin}, {Lissauer}, {MacQueen},
  {Marcy}, {McCullough}, {Morton}, {Narita}, {Paegert}, {Palle}, {Pepe},
  {Pepper}, {Quirrenbach}, {Rinehart}, {Sasselov}, {Sato}, {Seager},
  {Sozzetti}, {Stassun}, {Sullivan}, {Szentgyorgyi}, {Torres}, {Udry}, \&
  {Villasenor}}]{ricker14}
{Ricker}, G.~R., {Winn}, J.~N., {Vanderspek}, R., {et~al.} 2014, in \procspie,
  Vol. 9143, Space Telescopes and Instrumentation 2014: Optical, Infrared, and
  Millimeter Wave, 914320

\bibitem[{Rousseeuw \& Croux(1993)}]{rousseeuw93}
Rousseeuw, P.~J., \& Croux, C. 1993, Journal of the American Statistical
  Association, 88, 1273

\bibitem[{{Seager} \& {Mall{\'e}n-Ornelas}(2003)}]{seager03}
{Seager}, S., \& {Mall{\'e}n-Ornelas}, G. 2003, \apj, 585, 1038

\bibitem[{{Shabram} {et~al.}(2016){Shabram}, {Demory}, {Cisewski}, {Ford}, \&
  {Rogers}}]{shabram16}
{Shabram}, M., {Demory}, B.-O., {Cisewski}, J., {Ford}, E.~B., \& {Rogers}, L.
  2016, \apj, 820, 93

\bibitem[{{Shibahashi} \& {Kurtz}(2013)}]{shibahashi12}
{Shibahashi}, H., \& {Kurtz}, D.~W. 2013, in Astronomical Society of the
  Pacific Conference Series, Vol. 479, Progress in Physics of the Sun and
  Stars: A New Era in Helio- and Asteroseismology, ed. H.~{Shibahashi} \& A.~E.
  {Lynas-Gray}, 503

\bibitem[{{Sing}(2010)}]{sing10}
{Sing}, D.~K. 2010, \aap, 510, A21

\bibitem[{Smirnov(1948)}]{s48}
Smirnov, N. 1948, Ann. Math. Statist., 19, 279

\bibitem[{{Sullivan} {et~al.}(2015){Sullivan}, {Winn}, {Berta-Thompson},
  {Charbonneau}, {Deming}, {Dressing}, {Latham}, {Levine}, {McCullough},
  {Morton}, {Ricker}, {Vanderspek}, \& {Woods}}]{sullivan15}
{Sullivan}, P.~W., {Winn}, J.~N., {Berta-Thompson}, Z.~K., {et~al.} 2015, \apj,
  809, 77

\bibitem[{{Teachey} {et~al.}(2017){Teachey}, {Kipping}, \&
  {Schmitt}}]{teachey17}
{Teachey}, A., {Kipping}, D.~M., \& {Schmitt}, A.~R. 2017, ArXiv e-prints,
  arXiv:1707.08563

\bibitem[{{Van Eylen} \& {Albrecht}(2015)}]{vea15}
{Van Eylen}, V., \& {Albrecht}, S. 2015, \apj, 808, 126

\bibitem[{{Wang} \& {Ford}(2011)}]{wang11}
{Wang}, J., \& {Ford}, E.~B. 2011, \mnras, 418, 1822

\bibitem[{{Winn}(2010)}]{winn10}
{Winn}, J.~N. 2010, {Exoplanet Transits and Occultations}, ed. S.~{Seager},
  55--77

\bibitem[{{Yee} \& {Gaudi}(2008)}]{yee08}
{Yee}, J.~C., \& {Gaudi}, B.~S. 2008, \apj, 688, 616

\end{thebibliography}

\end{document}